\documentclass[reprint,
groupedaddress,
unsortedaddress,
runinaddress,
frontmatterverbose, 
nofootinbib,
nobibnotes,
amsmath,amssymb,
aps,
pra,
superscriptaddress,
]{revtex4-2}
\usepackage[dvipsnames]{xcolor}
\usepackage{graphicx,lipsum}
\usepackage{subcaption}
\usepackage{hyperref}
\usepackage{nameref}
\usepackage[justification=raggedright,singlelinecheck=false]{caption}
\usepackage[labelfont=bf, font=small]{caption}
\usepackage[font=footnotesize]{caption}
\usepackage{tikz}
\usetikzlibrary{arrows.meta, positioning, shapes}

\usepackage{float}
\usepackage{amsthm}
\usepackage{amsmath}

\usepackage{booktabs}
\usepackage{algorithm}
\usepackage{pgfplots}
\pgfplotsset{compat=1.18}
\usepackage{algpseudocode}
\usepackage{amssymb}
\usepackage{braket}
\usepackage{stackengine}
\usepackage{bm}
\usepackage[utf8]{inputenc}
\usepackage{lmodern}
\usepackage{mathtools}
\usepackage{tabularray}
\usepackage[mathlines]{lineno}
\usepackage{fancyhdr}
\usepackage{babel}
\usepackage{csquotes}
\usepackage{natbib}

\begin{document}

\preprint{APS/123-QED}

\title{Dynamics and Control of Two Coupled Quantum Oscillators: An Analytical Approach}


\author{Ali Abu-Nada}
\affiliation{Sharjah Maritime Academy, Sharjah, United Arab Emirates}
\author{Lian-Ao Wu}
\affiliation{Department of Physics, University of the Basque Country UPV/EHU, 48080 Bilbao, Spain}
\affiliation{IKERBASQUE, Basque Foundation for Science, Bilbao 48011, Spain}
\affiliation{EHU Quantum Center, University of the Basque Country UPV/EHU, Leioa, Biscay 48940, Spain}

\begin{abstract}
We analyze two coupled quantum oscillators in a common Lorentzian environment and control them by detuning (temporarily shifting) their frequencies. The reduced dynamics are solved exactly, without Born or Markov approximations, by propagating each detuning segment in closed form. We study two control schedules: \emph{regular} detuning, with perfectly periodic on and off pulses of fixed period, width, and amplitude; and \emph{irregular} detuning, with the same on/off structure but cycle-to-cycle jitter in period, width, and/or amplitude. Our main observable is the average excitation number (AEN) of each mode. Detuning moves the system away from the bath’s spectral peak, suppressing decoherence and damping non-Markovian revivals; in effectively Markovian baths the benefit is small. We quantify performance with a simple time-domain suppression factor. Larger detuning amplitudes and higher duty cycles yield stronger protection. Irregular control is slightly weaker at low duty cycle but becomes comparable to regular control as the duty cycle approaches one. These results give practical design rules linking detuning, duty cycle, and bath width, and provide an exact benchmark for controlled non-Markovian dynamics.
\end{abstract}

\maketitle
\section{Introduction}\label{sec:sec1}

Open quantum systems are fundamental in many areas of physics, including quantum optics, condensed matter, and quantum information science~\cite{nielsen, breuer2002theory, rivas2, rivas}. In realistic models, the system of interest interacts with its surrounding environment, typically described as having infinitely many degrees of freedom. The reduced dynamics of the system are then found by tracing out the environment’s degrees of freedom~\cite{nielsen2010quantum, banerjee2018open, breuer2002theory, rivas}.

To make this problem tractable, most standard approaches rely on two major approximations: the Born approximation, which assumes weak system--environment coupling and that the environment remains effectively unchanged by the system dynamics~\cite{breuer2002theory, Weiss2012, GardinerZoller2004}, and the Markov approximation, which neglects any memory effects in the bath~\cite{breuer2002theory, breuer, Gaikwad}. These approximations lead to widely used frameworks such as the Lindblad-Markovian master equation and the Redfield equation~\cite{Lindblad1976, Redfield1957}, which remain foundational tools in quantum optics~\cite{scully}, condensed matter physics~\cite{Mahan2000}, and quantum information science~\cite{nielsen2010quantum}.

However, these approximations break down when the system is strongly coupled to its environment, when the environment retains significant memory, or when the bath’s spectral features and initial correlations cannot be neglected. In such non-Markovian regimes \cite{abunada,PhysRevA.110.052209}, the environment can feed back information or energy to the system, resulting in revivals, backflow, and more complex dynamics that go beyond the reach of simple Markovian models. Exact solutions that fully capture non-Markovian effects are rare but provide critical insight into the true behavior of realistic open quantum systems.

Recently, a notable contribution by Wang \emph{et~al.}~\cite{wu} demonstrated how an exactly solvable open quantum system can reveal both Markovian and non-Markovian dynamics without relying on these standard approximations. Specifically, they derived an exact analytical solution for the dynamics of a single bosonic oscillator coupled to a Lorentzian reservoir, showing clearly how finite bath memory produces relaxation, energy revivals, and backflow that simple Markovian models fail to capture.

In parallel, Wu~\emph{et~al.} studied two non-interacting harmonic oscillators coupled simultaneously to a common environment, introducing a tunable parameter that interpolates between local and collective decoherence. Working in the Heisenberg picture, they obtained exact operator dynamics and characterized conditions under which \emph{steady-state quantum coherence} (SSQC) appears or is suppressed in the \emph{partial collective, non-Markovian} regime~\cite{WuMaWangBrumerWu2025SSC}. Their analysis emphasizes how the balance of collective vs.\ individual decoherence controls long-time behavior and clarifies when non-Markovianity does \emph{not} by itself guarantee SSQC.

\begin{figure}[h]
    \centering
    \includegraphics[width=1.0\linewidth]{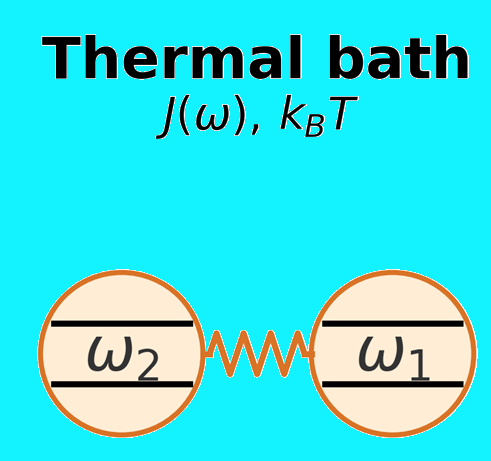}
    \caption{Schematic of the model: two bosonic oscillators with direct coupling, both interacting with a common thermal bath. The shared reservoir mediates additional indirect interactions and induces non-Markovian memory effects. In our analysis, dynamical decoupling (DD) is implemented as fast detuning of the mode frequencies to move the system off resonance with the bath’s spectral peak; inter-mode coherence is used only as a secondary diagnostic observable~\cite{mandel1995optical,walls2008quantum}.}
    \label{fig:fig1}
\end{figure}

Here we analyze a minimal yet richer model: two directly interacting bosonic oscillators that are both coupled to a common Lorentzian reservoir. We solve the controlled dynamics in closed form by propagating each detuning segment exactly and use the average excitation numbers as the primary figures of merit \textit{(see Fig.~\ref{fig:fig1})}. This framework naturally describes platforms such as coupled optical cavities~\cite{scully, yariv2000coupled} and superconducting circuits with a shared element~\cite{wendin2017quantum}, and it captures two pathways for energy exchange: direct mode--mode coupling through a coherent Hamiltonian interaction and indirect exchange mediated by the shared bath~\cite{braun2002creation}.

The exact solution of the system’s dynamics allows us to detect non-Markovian memory effects and quantify how the environment governs energy transfer between the modes. In our parameter regimes, strong revivals and significant energy exchange can arise from the reservoir’s finite memory and correlation time. In this sense, the bath acts as both a sink and a dominant channel for indirect backflow, making it possible to diagnose non-Markovian behavior beyond the direct link. Throughout, inter-mode coherence is used only as a \emph{diagnostic} of bath-mediated correlations; we do not target entanglement generation, and none of our conclusions relies on coherence as a primary figure of merit~\cite{mandel1995optical, walls2008quantum}.

Importantly, our approach naturally respects probability conservation and ensures that all energy and coherence transfer within the system--bath setup is fully accounted for by the exact solution itself. Because the entire environment is explicitly included in the model, any energy that leaves one oscillator enters the shared reservoir and can later flow back, depending on the bath’s memory. This means that the physical processes which would appear as an explicit “driving term” in approximate treatments are already fully described by the system’s internal dynamics and the reservoir’s finite correlations. In other words, there is no need to introduce or solve a separate inhomogeneous source term: the environment here is not an external classical driver but an integral part of the total quantum system, which evolves unitarily when the system and bath are treated together. When the reservoir is traced out to obtain the reduced system dynamics, any backflow, revival, or induced coherence appears naturally through the exact time evolution. This guarantees that our solution remains fully self-consistent: the total probability and energy are conserved at all times, no information is artificially lost or injected, and all indirect environment-mediated effects emerge directly from the finite memory of the bath. Unlike approximate Markovian models, which add dissipative or driving terms by hand, this exact approach preserves the complete dynamical structure and provides a rigorous way to detect when and how the reservoir’s memory generates observable non-Markovian effects (with coherence treated as a secondary observable) between initially uncorrelated, coupled quantum modes.

Using an exact model with no approximations, we ask a simple question: \emph{when do detuning pulses reduce decoherence in structured baths?}

\begin{figure}[h!]
    \centering
    \includegraphics[width=1.0\linewidth]{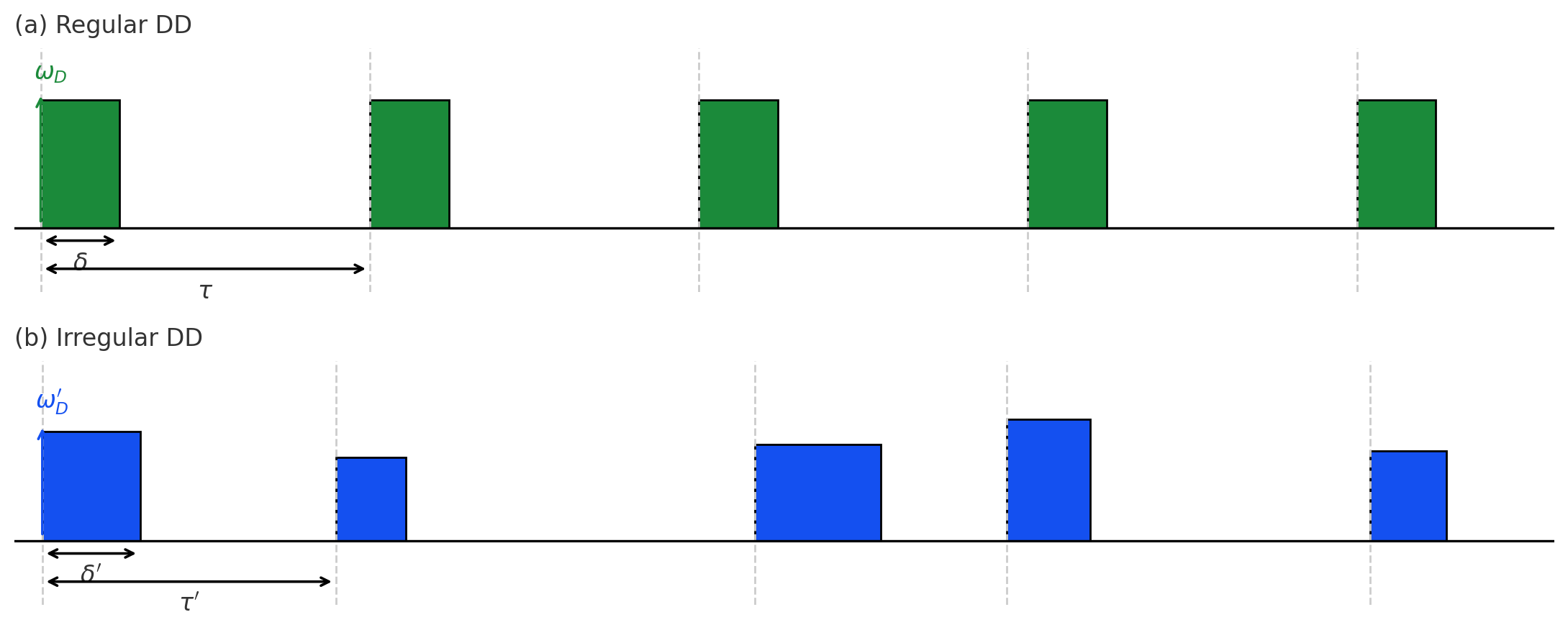}
    \caption{Regular vs.\ irregular detuning–based dynamical–decoupling (DD) pulse trains. 
(a) Regular (periodic) rectangular DD with fixed detuning amplitude $\omega_D$, pulse width $\delta$, and period $\tau$ (duty cycle $\eta=\delta/\tau$). 
(b) Irregular (randomized) DD in which the width $\delta_k$, period $\tau_k$, and detuning amplitude $\omega_{D,k}$ fluctuate from cycle to cycle (with $0<\delta_k<\tau_k$). 
Vertical dashed lines mark pulse-start times; double-headed arrows show the parameters on the first interval. 
Irregular DD models realistic timing/amplitude noise and is used to assess the robustness of the scheme.}
    \label{fig:fig2}
\end{figure}

In this work we adopt a \emph{detuning-based, leakage-elimination-operator (LEO)-inspired} dynamical decoupling (DD) scheme: rapid, coherent frequency shifts lift the system off resonance with the bath’s spectral peak and thereby reduce system--bath exchange~\cite{ViolaLloyd1998,ViolaKnillLloyd1999}. In the filter-function view, such frequency modulation acts as a spectral filter: by choosing the detuning magnitude and pulse timing, one suppresses noise where the bath carries weight~\cite{Cywinski2008,SuterAlvarez2016}. In short, for a memoryless (Markovian) bath there is little to recover and modulation helps only weakly; for a long-memory (non-Markovian) bath, good timing and sufficient detuning can strongly block information and energy from flowing back.\\
DD originated in NMR (spin echoes, CPMG) and was formalized for quantum systems as bang-bang decoupling~\cite{ViolaLloyd1998,ViolaKnillLloyd1999}. Many sequences have since been developed, including concatenated DD (robustness via hierarchy)~\cite{KhodjastehLidar2005} and Uhrig DD (spectrally targeted with optimized timing for certain baths)~\cite{Uhrig2007}. The effectiveness of DD has been demonstrated across platforms—trapped ions~\cite{Biercuk2009}, NV centers in diamond~\cite{deLange2010}, and superconducting qubits~\cite{Bylander2011}—and is now a standard primitive in quantum control.\\
Real devices, however, seldom realize perfectly periodic control. To model realistic imperfections, \emph{randomized/irregular DD} perturbs pulse widths, periods, and detuning amplitudes from cycle to cycle~\cite{ViolaKnill2005,SantosViola2006,KernAlber2005,Jing2014}. Such irregularity can average out coherent error buildup and increase robustness, but may also reduce peak suppression if the duty cycle is low. We therefore study both \emph{regular} DD (fixed period/width/detuning) and \emph{irregular} DD with $\pm 20\%$ jitter in all three parameters, following the nonperturbative framework of Ref.~\cite{Jing2014}. In our construction, DD enters as a fast, piecewise detuning of the oscillator frequencies—formally equivalent to a LEO-type frequency-modulation control—so that each ON/OFF segment is propagated exactly within the same homogeneous solution, and segments are matched by continuity of the amplitudes and their time derivatives (no phenomenology added). The timing parameters $(\tau,\delta,\eta,\omega_D)$ used throughout are illustrated in Fig.~\ref{fig:fig2}.

\noindent
In the non-Markovian regime (long bath memory), high-duty-cycle DD—each control \emph{cycle} of period $\tau$ with an ON window of duration $\delta$ (duty cycle $\eta\!\equiv\!\delta/\tau$) and detuning amplitude $\omega_D$ (the instantaneous frequency shift during the ON window; see Fig.~\ref{fig:fig2})—strongly suppresses heating and damps revivals by pushing the system off resonance with the bath’s Lorentzian peak at $\Omega$ and disrupting temporal correlations. By contrast, in the Markovian regime (short memory), DD yields little improvement: energy transferred to the reservoir is irreversibly dispersed and frequency modulation cannot recover it. These trends agree with the filter-function picture~\cite{Cywinski2008,SuterAlvarez2016} and prior non-Markovian analyses~\cite{Jing2014}.

\noindent
We quantify the impact of control with a single time-domain \emph{suppression factor} that compares the average excitation number (AEN) under control to the corresponding no-control trajectory. In our setting, growth of the AEN is a proxy for heating and coherence loss; thus, larger suppression indicates stronger mitigation of decoherence. Values close to one signify strong suppression, values near zero indicate little or no improvement, and negative values indicate degradation. For concise summaries we also report a window-averaged suppression factor over a fixed time interval.

\section{Model and Exact Mathematical Framework}\label{sec:sec2}

We study a minimal but nontrivial open system of two directly interacting bosonic modes that also share a common thermal reservoir. This extends exact single-mode resonance models to a two-mode, interacting setting and captures both direct mode--mode coupling and indirect, bath-mediated exchange within one exactly solvable bilinear framework. The total Hamiltonian reads:
\begin{equation} \label{eq:eq1}
    H = H_S + H_B + H_{SB},
\end{equation}
where the system Hamiltonian is
\begin{equation}\label{eq:eq2}
  H_S =  \omega_1\, a_1^\dagger a_1 + \omega_2\, a_2^\dagger a_2 + g\, ( a_1^\dagger a_2 + a_2^\dagger a_1 ).
\end{equation}
Here $a_\ell$ ($\ell=1,2$) is the bosonic \emph{annihilation} operator for mode $\ell$, and 
$a_\ell^\dagger$ is its Hermitian-conjugate \emph{creation} operator. They satisfy the canonical
commutation relations $[a_\ell, a_m^\dagger]=\delta_{\ell m}$, $[a_\ell, a_m]=[a_\ell^\dagger, a_m^\dagger]=0$,
and the number operator $n_\ell \equiv a_\ell^\dagger a_\ell$ counts the excitations in mode $\ell$.
We set $\hbar=1$ throughout. The bilinear term $g(a_1^\dagger a_2 + a_2^\dagger a_1)$ is a
beam-splitter (coherent hopping) interaction that exchanges quanta between the two modes.
The parameters $\omega_1$ and $\omega_2$ are the natural (bare) frequencies of the modes, and
$g$ is their direct coherent coupling. The bath Hamiltonian is
\begin{equation}
H_B = \sum_{j} \omega_j\, b_j^\dagger b_j,
\label{eq:eq3}
\end{equation}
where $j$ indexes the harmonic modes of the bath (reservoir) of frequency $\omega_j$, and
$b_j$ ($b_j^\dagger$) are bosonic \emph{annihilation}/\emph{creation} operators obeying $[\,b_j, b_k^\dagger\,]=\delta_{jk}$ and $[\,b_j,b_k\,]=[\,b_j^\dagger,b_k^\dagger\,]=0$.
The number operator $n_j \equiv b_j^\dagger b_j$ counts excitations in mode $j$.
In thermal equilibrium at temperature $T_B$, the mean occupation of the mode $j$ is $\langle n_j\rangle = \big(e^{\beta \omega_j}-1\big)^{-1}$ and $\beta=\left(k_{B} T_{B}\right)^{-1}$, with $k_B$ the Boltzmann constant. The interaction Hamiltonian is
\begin{equation}
H_{SB} = \sum_{j}\!\left[\, \kappa_{1j}\!\left(a_1^\dagger b_j + b_j^\dagger a_1\right)
+ \kappa_{2j}\!\left(a_2^\dagger b_j + b_j^\dagger a_2\right) \right],
\label{eq:eq4}
\end{equation}
where $\kappa_{1j},\kappa_{2j}\in\mathbb{C}$ are system--bath coupling constants that set the strength of
\emph{excitation exchange} between system mode $\ell$ and bath mode $j$.

\noindent
The Heisenberg equations of motion follow directly from the total Hamiltonian (see Appendix~\ref{appendixA}):
\begin{equation}
\begin{aligned}
\frac{d}{dt} a_1 &= -i\, \omega_1\, a_1 - i\, g\, a_2 - i \sum_j \kappa_{1j} b_j, \\
\frac{d}{dt} a_2 &= -i\, \omega_2\, a_2 - i\, g\, a_1 - i \sum_j \kappa_{2j} b_j, \\
\frac{d}{dt} b_j &= -i\, \omega_j\, b_j - i\, \kappa_{1j} a_1 - i\, \kappa_{2j} a_2.
\end{aligned}
\label{eq:eq5}
\end{equation}

Solving for \(b_j(t)\) gives:
\begin{align}
  b_j(t) &= b_j(0)\, e^{-i \omega_j t} 
  - i\, \kappa_{1j} \int_0^t ds\, a_1(s)\, e^{-i \omega_j (t-s)} \notag \\ 
  &\quad - i\, \kappa_{2j} \int_0^t ds\, a_2(s)\, e^{-i \omega_j (t-s)}.
  \label{eq:eq6}
\end{align}

Substituting Eq.~\ref{eq:eq6} into \ref{eq:eq5} yields:
\begin{align}
\dot{a}_1(t) &= -i\, \omega_1\, a_1(t) - i\, g\, a_2(t) + F_{1}(t) \notag \\
&\quad - \int_0^t ds\, G(t-s)\, [\, a_1(s) + a_2(s) ], \nonumber \\[8pt]
\dot{a}_2(t) &= -i\, \omega_2\, a_2(t) - i\, g\, a_1(t) + F_{2}(t) \notag \\
 &\quad - \int_0^t ds\,G(t-s)\, [\, a_1(s) + a_2(s) ].
\label{eq:eq7}
\end{align}
Assuming both system modes couple equally to the same bath, set \(\kappa_{1j} = \kappa_{2j} = \kappa_j\) for all \(j\). Then the noise terms for both modes are identical: \(F_1(t) = F_2(t) = F(t) = -i \sum_j \kappa_j\, b_j(0)\, e^{-i \omega_j t}\). Physically, this means the reservoir injects the same quantum noise into both oscillators, so any coherence or energy transfer that appears between them arises from both direct exchange and the bath’s finite memory.

The memory kernel
\begin{align}
G(t-s) &= \sum_j \kappa_j^2\, e^{-i \omega_j (t-s)} \notag \\
       &= \int d\omega\, J(\omega)\, e^{-i \omega (t-s)}, 
\label{eq:eq8}
\end{align}
where $J(\omega) = \frac{\Gamma \gamma}{2\pi}\,\frac{1}{(\omega - \Omega)^2 + \gamma^2}$ is the spectral density of the reservoir modes: the Lorentzian form specifies that the reservoir is peaked around a central frequency \(\Omega\) with width \(\gamma\). The parameter \(\Omega\) sets the resonance frequency of the bath, while \(\Gamma\) controls the overall coupling strength between the system and the environment. $G(t-s)$ encodes how the reservoir’s correlations influence the system’s dynamics at earlier times. A larger \(\Gamma\) indicates stronger system--bath interaction, while the inverse width \(\gamma^{-1}\) sets the reservoir’s correlation time: smaller \(\gamma\) implies longer memory. When \(G(t-s)\) remains non-zero for finite delays, the bath feeds information back into the system, producing non-Markovian effects such as revivals and backflow of energy and coherence. 

We adopt the resonance condition where both system modes are set equal to the reservoir’s central frequency: $\omega_{1} = \omega_{2} = \Omega$.
This ensures that the reservoir’s correlations are maximally relevant for the system’s dynamics, especially at earlier times.

When the bath correlation function $G(t - s)$ remains non-zero for finite delays, the reservoir can feed information back into the system, producing non-Markovian effects such as revivals and partial backflow of energy and coherence. In contrast, for large $\gamma$ (broad spectral width), the reservoir acts effectively memoryless, yielding purely Markovian decay.

We expand:
\begin{equation}
    a_1(t) = A_1(t) a_1(0) + A_2(t) a_2(0) + \sum_j B_{1j}(t) b_j(0),
    \label{eq:eq9}
\end{equation}
\begin{equation}
a_2(t) = C_1(t) a_1(0) + C_2(t) a_2(0) + \sum_j B_{2j}(t) b_j(0).
\label{eq:eq10}
\end{equation}

By symmetry, the direct coupling term and the shared bath treat both modes in the same way: each mode can exchange energy with the other through the direct hopping \( g \), and both modes interact identically with the same reservoir. Physically, this means that any excitation initially in mode~1 can flow into mode~2 in exactly the same way that an excitation initially in mode~2 can flow back into mode~1. Mathematically, this symmetry implies $C_1(t) = A_2(t)$ and $C_2(t) = A_1(t)$.

\noindent
To find the linear, first-order ordinary differential equations for the amplitudes, substitute the operator expansions for $a_1(t)$ and $a_2(t)$ in Eq.~\ref{eq:eq9} and Eq.~\ref{eq:eq10} back into the Heisenberg equations of motion, then equate coefficients of the linearly independent initial operators $a_1(0)$, $a_2(0)$, and $b_j(0)$. This yields a closed system of coupled integro-differential equations for the mode amplitudes and the bath amplitudes:
\[
\begin{aligned}
\dot{A}_1(t) &= -i\, \omega_1\, A_1(t) - i\, g\, A_2(t) 
              - \int_0^t ds\, G(t-s)\, [\, A_1(s) + A_2(s) ], \\[6pt]
\dot{A}_2(t) &= -i\, \omega_2\, A_2(t) - i\, g\, A_1(t) 
              - \int_0^t ds\, G(t-s)\, [\, A_1(s) + A_2(s) ].
\end{aligned}
\]

Physically, these equations show that the rate of change of each mode's amplitude depends on three effects:
(1) its own natural frequency,
(2) coherent exchange with the other mode through the direct coupling $g$,
and (3) the memory effect of the shared bath, captured by the integral term with $G(t-s)$.

Similarly, matching the coefficients of each bath operator $b_j(0)$ yields the equations for the bath amplitudes:
\[
\begin{aligned}
\dot{B}_{1j}(t) &= -i\, \omega_j\, B_{1j}(t) - i\, \kappa_{1j}\, A_1(t) - i\, \kappa_{2j}\, A_2(t), \\[6pt]
\dot{B}_{2j}(t) &= -i\, \omega_j\, B_{2j}(t) - i\, \kappa_{1j}\, A_1(t) - i\, \kappa_{2j}\, A_2(t).
\end{aligned}
\]

For the Lorentzian spectral density, the memory kernel is
\[
G(t - s) = \frac{\Gamma \gamma}{2}\, e^{-[\, \gamma + i \Omega\,]\, |t - s|},
\]
which satisfies
\[
\dot{G}(t - s) = -(\gamma + i \Omega)\, G(t - s).
\]

Differentiating yields explicit second-order ODEs for the mode amplitudes (homogeneous) together with inhomogeneous second-order ODEs for the bath amplitudes driven by the system modes:
\begin{align}
\ddot{A}_1 + \alpha\, \dot{A}_1 + \beta\, A_1 + \gamma\, A_2 &= 0, \label{eq:eq11} \\[6pt]
\ddot{A}_2 + \tilde{\alpha}\, \dot{A}_2 + \tilde{\beta}\, A_2 + \tilde{\gamma}\, A_1 &= 0, \label{eq:eq12}
\end{align}
where $\alpha = \gamma + i(\Omega + \omega_1)$, $\tilde{\alpha} = \gamma + i(\Omega + \omega_2)$, $\beta = \frac{\Gamma \gamma}{2} - \omega_1 \Omega + i \gamma \omega_1$, $\tilde{\beta} = \frac{\Gamma \gamma}{2} - \omega_2 \Omega + i \gamma \omega_2$, and $\tilde{\gamma} = i g(\gamma + i \Omega)$.

The bath amplitudes obey inhomogeneous second-order ODEs:
\begin{align}
\ddot{B}_{1j} + \alpha_B\, \dot{B}_{1j} + \beta_B\, B_{1j}
&= -\kappa_j\,[\,i \alpha + \omega_j]\, e^{-i \omega_j t}, \label{eq:eq13} \\[6pt]
\ddot{B}_{2j} + \alpha_B\, \dot{B}_{2j} + \beta_B\, B_{2j}
&= -\kappa_j\,[\,i \alpha + \omega_j]\, e^{-i \omega_j t}, \label{eq:eq14}
\end{align}
where $\alpha_B = \alpha + i \omega_j$ and $\beta_B = \beta - i\, \omega_j$.

The explicit bath terms sum to the missing probability:
\[
|A_1(t)|^2 + |A_2(t)|^2 + \sum_j |B_{1j}(t)|^2 = 1.
\]

We assume that at time $t = 0$, the total system is prepared in a product state where the two oscillators are in a pure quantum state and the reservoir is in thermal equilibrium. Specifically, the system state is $| \psi_0 \rangle = |1\rangle_1 \otimes |0\rangle_2$, meaning that mode~1 starts in the single-excitation Fock state and mode~2 starts in its vacuum state. The bath is assumed initially thermal, described by a product of thermal states for each bath oscillator mode, $
\rho_B = \bigotimes_j \rho_j, 
\quad
\rho_j = \frac{1}{Z_j} \sum_{n_j} e^{-\beta\, n_j\, \omega_j}\, | n_j \rangle \langle n_j |$, where $
Z_j = \sum_{n_j=0}^\infty e^{-\beta\, n_j\, \omega_j}$. The mean thermal occupation for each bath mode is $ n_B = \left(e^{\beta \omega_j} - 1\right)^{-1}$. Because the system and bath are uncorrelated at $t = 0$, the full initial density operator is $\rho(0) = | \psi_0 \rangle \langle \psi_0 | \otimes \rho_B$. 

Using the initial populations $n_{10} = \langle a_1^\dagger(0) a_1(0) \rangle = 1$, $n_{20} = \langle a_2^\dagger(0) a_2(0) \rangle = 0$, and $
n_B = \big\langle b_j^\dagger(0)\, b_j(0) \big\rangle 
= \left(e^{\beta \omega} - 1\right)^{-1}$,
the exact average excitation numbers (AENs) for the two modes are:
\begin{align}
\big\langle a_1^\dagger(t)\, a_1(t) \big\rangle 
&= |A_1(t)|^2\, n_{10} + |A_2(t)|^2\, n_{20} \notag \\
&\quad + \sum_j |B_{1j}(t)|^2\, n_B, 
\label{eq:eq15} \\[8pt]
\big\langle a_2^\dagger(t)\, a_2(t) \big\rangle 
&= |A_2(t)|^2\, n_{10} + |A_1(t)|^2\, n_{20} \notag\\
&\quad + \sum_j |B_{2j}(t)|^2\, n_B.
\label{eq:eq16}
\end{align}

Because the total probability is conserved in this exactly solvable model, the sum over the squared amplitudes satisfies
\[
|A_1(t)|^2 + |A_2(t)|^2 + \sum_j |B_{ij}(t)|^2 = 1,
\quad
(i = 1,2).
\]
This ensures that any energy leaving the system modes is fully accounted for by the bath.

Combining the explicit amplitudes with probability conservation, the exact AENs can be written as:
\begin{align}
n_{1}(t)=\big\langle a_1^\dagger(t)\, a_1(t) \big\rangle 
&= |A_1(t)|^2\, n_{10} + |A_2(t)|^2\, n_{20} \notag \\
&\quad + \big[\,1 - |A_1(t)|^2 - |A_2(t)|^2\,\big]\, n_B, 
\label{eq:eq17} \\[6pt]
n_{2}(t)=\big\langle a_2^\dagger(t)\, a_2(t) \big\rangle 
&= |A_2(t)|^2\, n_{10} + |A_1(t)|^2\, n_{20} \notag \\
&\quad + \big[\,1 - |A_1(t)|^2 - |A_2(t)|^2\,\big]\, n_B.
\label{eq:eq18}
\end{align}

Here, the final term captures the indirect contribution from the bath, which exactly accounts for the probability that flows out of the system modes and into the reservoir.

The inter-mode coherence between the two oscillators is given by the off-diagonal correlation function
\begin{align}
\big\langle a_1^\dagger(t)\, a_2(t) \big\rangle 
&= A_1^*(t)\, A_2(t)\, n_{10} + A_2^*(t)\, A_1(t)\, n_{20}  \notag \\
&\quad + \sum_j B_{1j}^*(t)\, B_{2j}(t)\, n_B.
\label{eq:eq19}
\end{align}
For our initial conditions, this reduces to
\[
\big\langle a_1^\dagger(t)\, a_2(t) \big\rangle 
= A_1^*(t)\, A_2(t) + \sum_j |B_{1j}(t)|^2\, n_B,
\]
so under symmetry the bath-induced part of the coherence is purely real and directly linked to the total probability that leaks into the environment. Combining with probability conservation, the exact inter-mode coherence becomes:
\begin{align}
n_{12}(t)= \big\langle a_1^\dagger(t)\, a_2(t) \big\rangle 
&= A_1^*(t)\, A_2(t) \notag \\ &\quad + \big[\,1 - |A_1(t)|^2 - |A_2(t)|^2\,\big]\, n_B.
\label{eq:eq20}
\end{align}
A non-zero value indicates phase coherence induced by direct coupling or by the shared reservoir’s memory. However, non-zero coherence does not by itself imply genuine quantum entanglement. A sufficient condition for entanglement is that the inter-mode coherence violates the Cauchy--Schwarz inequality~\cite{mandel1995optical, walls2008quantum}:
\begin{align}
|\big\langle a_1^\dagger(t)\, a_2(t) \big\rangle|^2 
> \big\langle a_1^\dagger(t)\, a_1(t) \big\rangle\, 
\big\langle a_2^\dagger(t)\, a_2(t) \big\rangle. 
\label{eq:eq21}
\end{align}
If this inequality is not satisfied, the coherence can be explained by classical correlations alone.

The final closed forms of Eqs.~(\ref{eq:eq17}) and (\ref{eq:eq18}) are provided in Appendix~\ref{AppendixB}, where we solve the homogeneous second-order linear differential equations explicitly to obtain $A_1(t)$ and $A_2(t)$.

\section{Dynamical Decoupling Control in the Non-Markovian Two-Oscillator System}
\label{sec:sec3}

We extend our exact analytical solution by adding an external \emph{dynamical decoupling} (DD) field to mitigate decoherence from a structured thermal reservoir~\cite{WuLidar2002,WuLidar2004}. Throughout this work we employ a \emph{LEO-inspired, detuning (frequency-modulation) DD}: we modulate the system \emph{transition frequencies} to suppress the system--bath coupling. This follows the Leakage--Elimination--Operator (LEO) idea~\cite{WuByrdLidar2002,ByrdLidarWuZanardi2005}, where fast coherent controls symmetrize the dynamics so that unwanted couplings average (nearly) to zero in an \emph{open-loop} manner without measurements~\cite{ViolaLloyd1998,ViolaKnillLloyd1999,SuterAlvarez2016,scully}. For completeness we note the alternative \emph{inversion ($\pi$-pulse) DD} picture (spin–echo/CPMG/UDD) and its filter-function interpretation~\cite{HaeberlenWaugh1968,Haeberlen1976,ErnstBodenhausenWokaun,Magnus1954,Cywinski2008,SuterAlvarez2016,KhodjastehLidar2005,Uhrig2007}, but we \emph{do not} use $\pi$ pulses here.

\paragraph*{Control law (LEO-inspired detuning).}
We apply a piecewise-constant detuning to both modes,
\begin{equation}
f_{\text{DD}}(t)=
\begin{cases}
\omega_D, & n\tau < t < n\tau + \delta,\\[4pt]
0, & \text{otherwise},
\end{cases}
\qquad n=0,1,2,\ldots,
\label{eq:eq22}
\end{equation}
with amplitude $\omega_D$, width $\delta$, period $\tau$, and duty cycle $\eta=\delta/\tau$. The control Hamiltonian
\begin{equation}
H_c(t)=f_{\mathrm{DD}}(t)\,\big(a_1^\dagger a_1+a_2^\dagger a_2\big)
\label{eq:eq23}
\end{equation}
shifts the instantaneous mode frequencies to
\begin{equation}
\omega_{i,\text{eff}}(t)=\omega_i+f_{\text{DD}}(t),\quad i=1,2,
\label{eq:eq24}
\end{equation}
so that during ON intervals the interaction acquires a fast phase $e^{i\phi(t)}$ with $\phi(t)=\!\int^t f_{\text{DD}}(t')\,dt'$. The associated filter function
\begin{equation}
F(\omega)=\Big|\!\int_0^T e^{i[\omega t+\phi(t)]}\,dt\Big|^2,
\label{eq:eq25}
\end{equation}
develops spectral notches near dominant bath frequencies~\cite{Cywinski2008,SuterAlvarez2016}; the limit $\eta\!\to\!1$ approaches near-continuous modulation. Operationally, this acts as a LEO in the energy basis: rapid gap modulation suppresses leakage/noise channels~\cite{WuByrdLidar2002,ByrdLidarWuZanardi2005}.

\noindent
The DD field partitions the evolution into alternating ON/OFF segments. In each segment, the homogeneous second-order ODEs for the mode amplitudes [Eqs.~(\ref{eq:eq11})–(\ref{eq:eq12})] retain their form with
\[
\text{ON: }\;\omega_i\mapsto\omega_{i,\mathrm{eff}}=\omega_i+\omega_D, \qquad
\text{OFF: }\;\omega_{i,\mathrm{eff}}=\omega_i.
\]
Equivalently, in the Laplace picture the characteristic quartic keeps its structure; when DD is ON, the resonance $\omega_{1,2}=\Omega$ is lifted by $\omega_D$, and the quartic in App.~\ref{AppendixB} reads
\begin{equation}
\lambda^4 + A\, \lambda^3 + B\, \lambda^2 + C\, \lambda + D = 0,
\label{eq:eq26}
\end{equation}
with shifted coefficients $\alpha_D,\tilde{\alpha}_D,\beta_D,\tilde{\beta}_D$ listed in App.~\ref{AppendixB}. We solve each segment exactly and match amplitudes and first derivatives at every switch, yielding a fully analytic, piecewise-exact evolution (no phenomenology).

\noindent
Detuning repeatedly lifts resonance with the Lorentzian bath peak at $\Omega$, reducing spectral overlap and energy exchange. The amplitude $\omega_D$ sets \emph{how far} we move off resonance; the duty cycle $\eta$ sets \emph{how long} protection is applied each period. Large $\omega_D$ and $\eta\!\to\!1$ maximize suppression of heating and non-Markovian revivals, consistent with toggling-frame and filter-function intuition~\cite{Cywinski2008,SuterAlvarez2016,Jing2014}. In effectively Markovian baths ($\gamma > \Gamma$), DD has little impact because dissipated energy is not recovered~\cite{Jing2014}.

\subsection{Regular vs.\ irregular DD}
\label{sec:dd_regular_irregular}
We consider two implementations of the same detuning-based (LEO-inspired) DD. \emph{Regular} DD uses the perfectly periodic square wave of Eq.~(\ref{eq:eq22}). \emph{Irregular} DD keeps that structure but perturbs each cycle to model realistic control noise and avoid coherent error build-up:
\begin{equation}
X_k = X + D_X\,\xi_k,\qquad X\in\{\delta,\tau,\omega_D\},\quad \xi_k\sim \mathrm{Unif}[-1,1],
\label{eq:eq27}
\end{equation}
subject to $0<\delta_k<\tau_k$ each cycle. Randomized decoupling is known to break error synchronization and improve robustness to device drifts and model uncertainty~\cite{ViolaKnill2005,SantosViola2006,KernAlber2005}; for long-memory baths, nonperturbative analyses report similar benefits~\cite{Jing2014}. The trade-off is that notches are \emph{smeared} (shallower/broader), so peak suppression is typically weaker than with perfectly periodic control.

\noindent
In long-memory baths, strong detuning and a high duty cycle are the primary levers for suppression: once the detuning amplitude $\omega_D$ and duty cycle $\eta$ are large, cycle-to-cycle irregularity has little practical downside. By contrast, at small $\eta$ the same irregularity shortens the time spent near the optimal notch frequencies of the control filter and can noticeably degrade performance. In effectively memoryless (Markovian) environments, $\gamma \gg \Gamma$, both regular and irregular DD provide little benefit.

\noindent
Implementation is identical for both schemes: we propagate ON/OFF segments with the same closed-form solutions [Eqs.~(\ref{eq:eq17})–(\ref{eq:eq18})], using $\omega_{i,\mathrm{eff}}=\omega_i+\omega_D$ (ON) and $\omega_{i,\mathrm{eff}}=\omega_i$ (OFF), and enforce continuity at every switch. For irregular DD we draw $\{\delta_k,\tau_k,\omega_{D,k}\}$ with $\pm20\%$ jitter (as in~\cite{Jing2014}) while maintaining $0<\delta_k<\tau_k$. This keeps the treatment analytic and enables a fair comparison of \emph{regular} vs.\ \emph{irregular} suppression across bath parameters $(\Gamma,\gamma,\Omega,T_B)$ and control knobs $(\omega_D,\eta)$.

\section{Results and Discussion}
\label{sec:results}

To demonstrate the exact non-Markovian dynamics of two oscillators interacting with a common thermal reservoir, we examine the time evolution of the average excitation numbers (AENs) $n_1(t)$ and $n_2(t)$ from Eqs.~(\ref{eq:eq17})–(\ref{eq:eq18}) across complementary regimes and controls. Unless stated otherwise, the initial state is $n_1(0)=1$ (single-excitation Fock state) and $n_2(0)=0$, and we work at resonance $\omega_1=\omega_2=\Omega$ to maximize bath-induced correlations.

\begin{figure*}[t]
    \centering
\begin{minipage}[t]{0.48\textwidth}
\centering
\includegraphics[width=\linewidth]{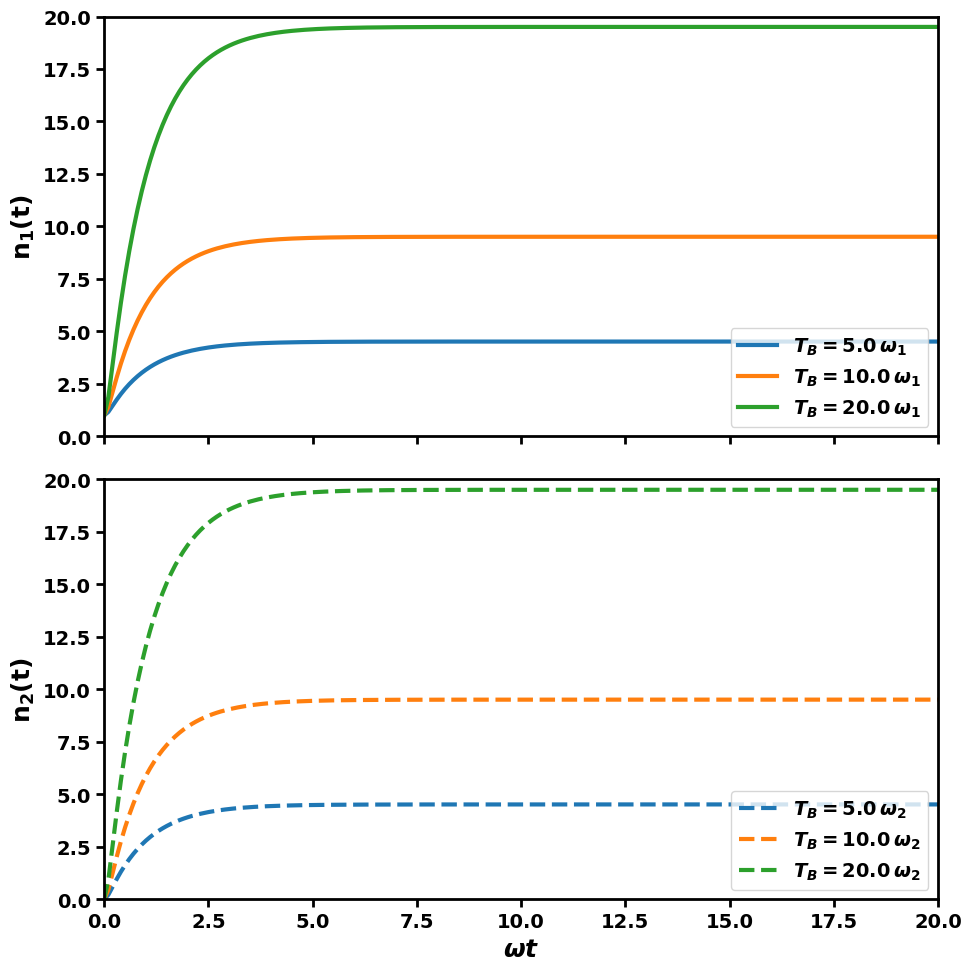} 
\caption{Time evolution of the mode occupation numbers in the Markovian regime. 
The upper panel shows $n_1(t)$ versus $\omega_1 t$ and the lower panel shows $n_2(t)$ versus $\omega_2 t$ 
for different bath temperatures $T_B$ (in units of $\omega_1$ and $\omega_2$, respectively). 
Parameters: $\Gamma = 1.0$, $\Omega = 1.0$, 
$\omega_1 = \omega_2 = 1.0$, and $\gamma = 15.0$. The large spectral width with $\gamma > \Gamma$ 
ensures a short reservoir correlation time, so the bath acts as a memoryless sink, 
leading to smooth relaxation without significant revivals or backflow.}
 \label{fig:fig3}
\end{minipage}
 \hfill
\begin{minipage}[t]{0.48\textwidth}
\centering
\includegraphics[width=\linewidth]{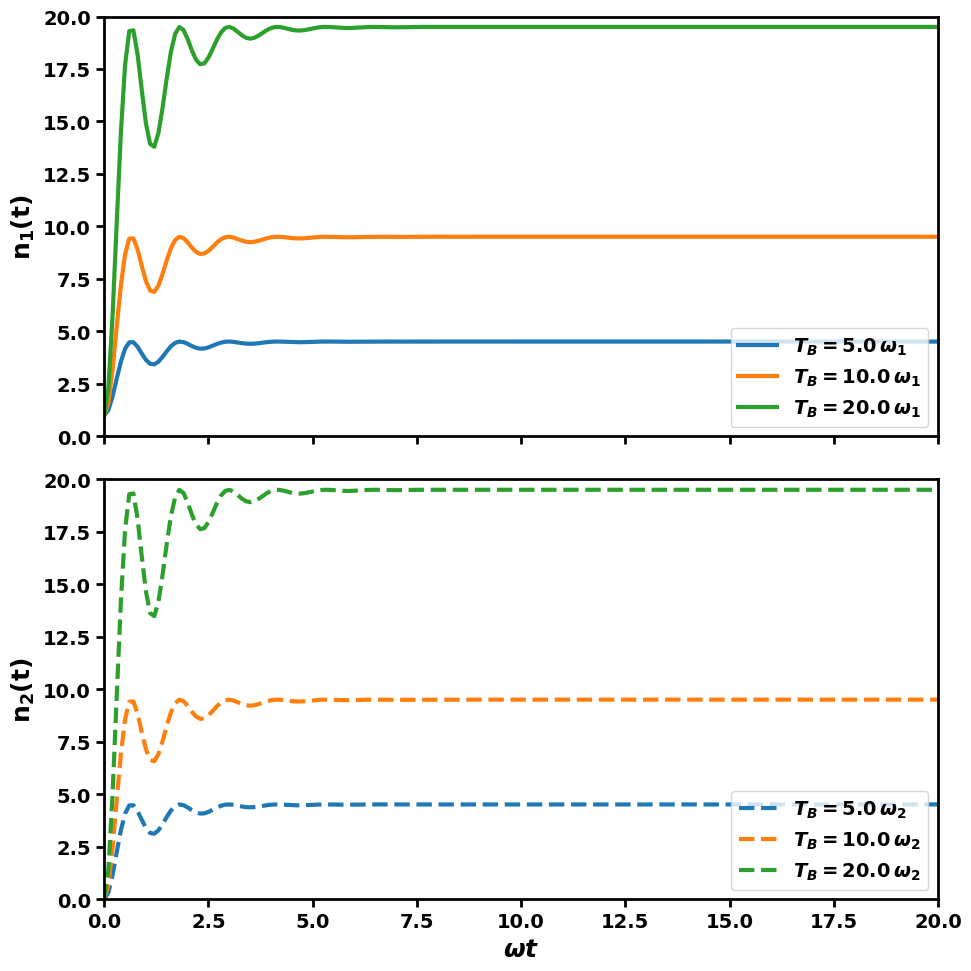} 
\caption{Time evolution of the mode occupation numbers in the non-Markovian regime. 
The upper panel shows $n_1(t)$ versus $\omega_1 t$ and the lower panel shows $n_2(t)$ versus $\omega_2 t$ 
for different bath temperatures $T_B$. 
Parameters: $\Gamma = 15.0$, $\Omega = 1.0$, 
$\omega_1 = \omega_2 = 1.0$, and $\gamma = 1.0$. Here, $\Gamma > \gamma$, resulting in a long reservoir correlation time and strong non-Markovian memory effects, which enable partial revivals and backflow of energy.}
\label{fig:fig4}
    \end{minipage}
\end{figure*}

\begin{figure}[h]
    \centering
    \includegraphics[width=1.0\linewidth]{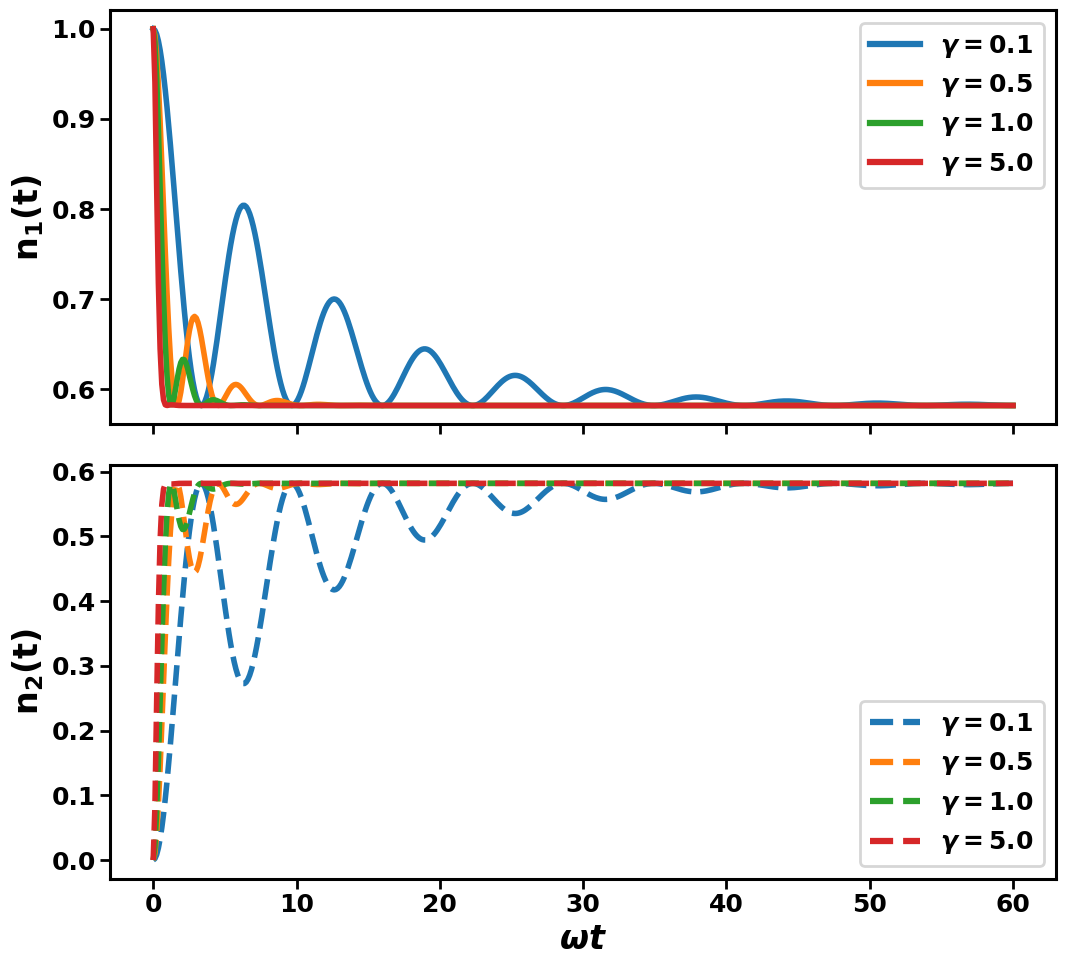}
   \caption{
Time evolution of the mode occupation numbers at fixed bath temperature $T_B=1.0$ for varying bath spectral widths $\gamma$. 
Top: $n_1(t)$ vs.\ $\omega t$; bottom: $n_2(t)$ vs.\ $\omega t$. 
Curves correspond to $\gamma=0.1$ (blue), $0.5$ (orange), $1.0$ (green), and $5.0$ (red). 
Parameters: $\Gamma=5.0$, $\Omega=1.0$, $\omega_1=\omega_2=1.0$. 
The bath correlation time scales as $\tau_c\!\sim\!1/\gamma$: small $\gamma$ (0.1, 0.5) produces pronounced non-Markovian oscillations and short-time overshoots, 
whereas as $\gamma$ increases to $1.0$ and $5.0$ the oscillation amplitude decreases and the approach becomes monotonic, indicating that a broadband reservoir quenches backflow. 
Both modes converge to the same thermal occupation $n_B$ set by $T_B$.}
\label{fig:fig5}
\end{figure}

\begin{figure}[h]
  \centering
  \begin{subfigure}{0.48\textwidth}
    \centering
    \includegraphics[width=\linewidth]{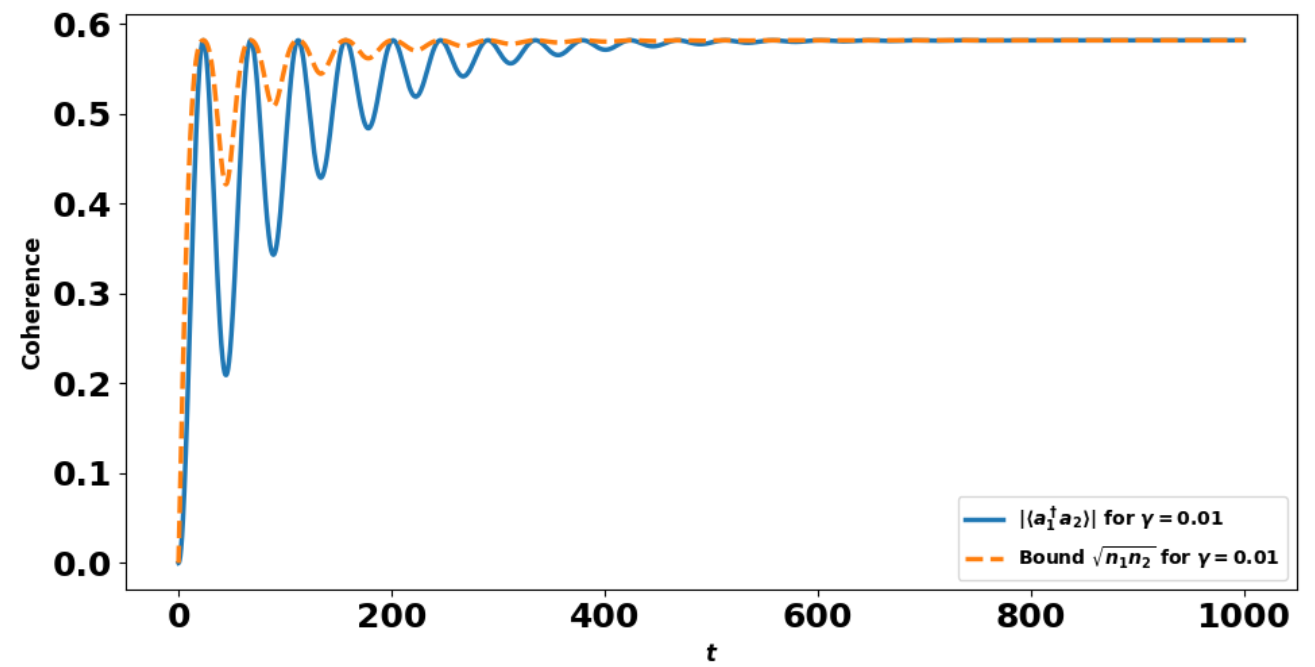}
    \caption{Non-Markovian regime (long memory): coherence exhibits oscillatory revivals and approaches—but never exceeds—the separability bound. 
    Parameters: $\Gamma=1.0$, $\gamma=0.01$, $\Omega=1.0$, $\omega_1=\omega_2=1.0$, $T_B=1.0$.}
    \label{fig:fig6a}
  \end{subfigure}\hfill
  \begin{subfigure}{0.48\textwidth}
    \centering
    \includegraphics[width=\linewidth]{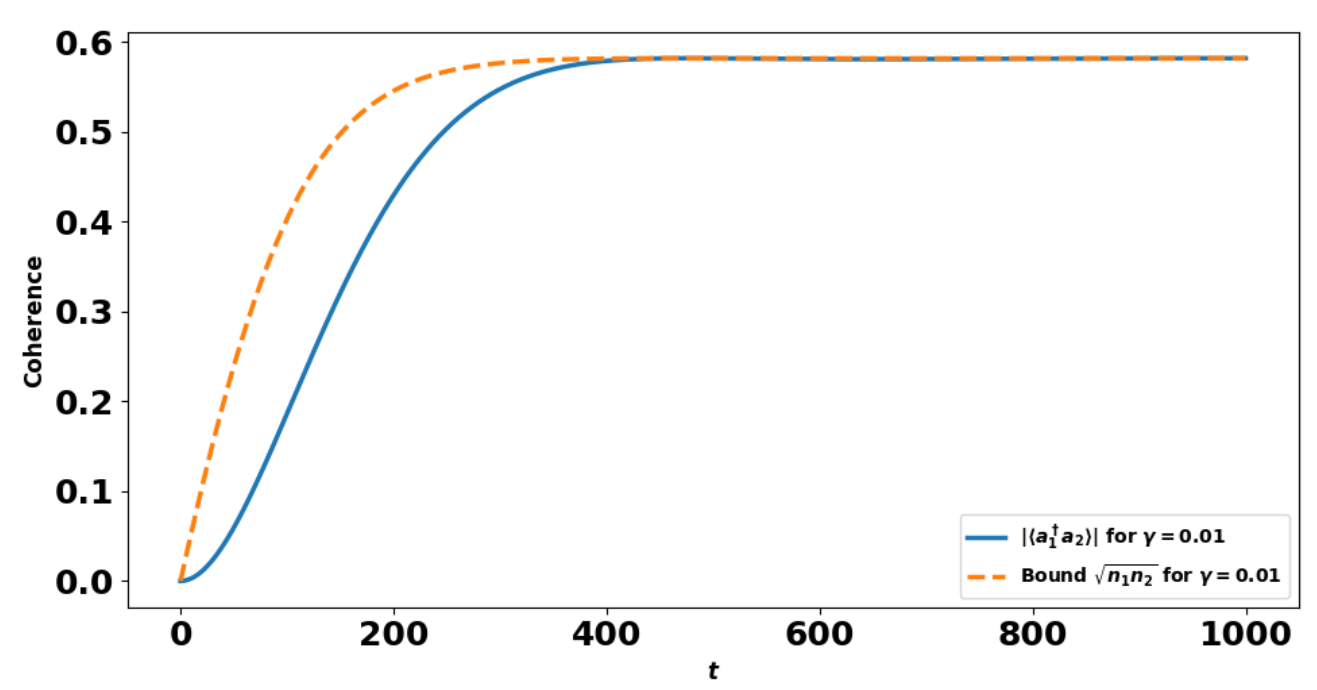}
    \caption{Markovian regime (short memory): coherence increases smoothly to a steady value below the separability bound, without revivals. 
    Parameters: $\Gamma=1.0$, $\gamma=15.0$, $\Omega=1.0$, $\omega_1=\omega_2=1.0$, $T_B=1.0$.}
    \label{fig:fig6b}
  \end{subfigure}
  \caption{Bath memory controls the coherence dynamics: revivals for long memory versus monotonic saturation for short memory; in both cases the separability limit is respected.}
  \label{fig:fig6}
\end{figure}

\subsection{Markovian vs.\ non-Markovian (Exact solution).}
In the Markovian case (Fig.~\ref{fig:fig3}; $\Gamma=1.0$, $\gamma=15.0$, $\Omega=\omega_1=\omega_2=1.0$), the broad bath ($\gamma\! > \!\Gamma$) yields a short correlation time $\tau_c\!\sim\!\gamma^{-1}$, so the reservoir is effectively memoryless. Energy exchange is one-way toward thermal equilibrium: for our initial conditions and $T_B$ used here, both $n_1(t)$ and $n_2(t)$ exhibit a smooth \emph{monotonic rise} (heating) from their initial values and asymptotically approach the common thermal occupation $n_B=(e^{\omega/T_B}-1)^{-1}$. Higher $T_B$ gives a larger asymptote, and no oscillations or revivals are observed. 

In contrast, in the non-Markovian regime (Fig.~\ref{fig:fig4}) we use 
$
\Gamma = 15.0,\quad \Omega = 1.0,\quad \omega_1 = \omega_2 = 1.0,\quad \gamma = 1.0.
$
Here the system--reservoir coupling $\Gamma$ is much larger than the spectral width $\gamma$ ($\Gamma > \gamma$), yielding a long reservoir correlation time that allows energy and information to flow back into the system. Consequently, the dynamics exhibit clear oscillations and partial revivals. In both cases, the final steady state is purely thermal and determined by the reservoir temperature, but the approach to equilibrium reveals whether the environment acts as a Markovian sink or a non-Markovian memory channel.

\paragraph*{Effect of bath spectral width on both modes.}
Figure~\ref{fig:fig5} shows $n_1(t)$ (top) and $n_2(t)$ (bottom) for four bath widths $\gamma\in\{0.1,0.5,1.0,5.0\}$ at fixed $T_B=1.0$. 
The spectral width sets the reservoir correlation time, $\tau_c\!\sim\!1/\gamma$. 
Small $\gamma$ (narrow spectrum) gives long $\tau_c$ and strong memory; large $\gamma$ (broad spectrum) gives short $\tau_c$ and effectively memoryless behavior. For mode 1 ($n_{1}(t)$), with $\gamma=0.1,0.5$ the dynamics are clearly non-Markovian: $n_1(t)$ displays underdamped oscillations and partial revivals before relaxing toward the thermal value $n_B$. 
As $\gamma$ increases to $1.0$ and $5.0$, the oscillation amplitude decreases and the evolution becomes monotonic, indicating that a broadband reservoir quenches backflow. For mode 2 ($n_{2}(t)$), starting from vacuum, $n_2(t)$ rises as excitations are transferred through the shared bath. 
For small $\gamma$ the rise is accompanied by pronounced oscillations and modest overshoots, out of phase with $n_1(t)$; for larger $\gamma$ the oscillations vanish and $n_2(t)$ increases smoothly and monotonically to the same thermal limit $n_B$ as $n_1(t)$.

\paragraph*{Coherence and separability.}
A shared bath can create coherence between the two modes even when direct coupling is present. 
With \emph{long} bath memory (non-Markovian, Fig.~\ref{fig:fig6a}), the coherence shows clear revivals and comes close to—but never crosses—the standard separability bound~\cite{mandel1995optical,walls2008quantum}. With \emph{short} memory (Markovian, Fig.~\ref{fig:fig6b}), backflow is suppressed and the coherence grows smoothly to a steady value below that bound. 
In short: memory sets \emph{revivals} vs.\ \emph{saturation}, and in both cases the coherence remains separable.

\subsection{DD control: amplitude vs.\ duty cycle (regular DD)}
\begin{figure*}[t]
\centering
\begin{subfigure}{0.48\textwidth}
  \includegraphics[width=\textwidth]{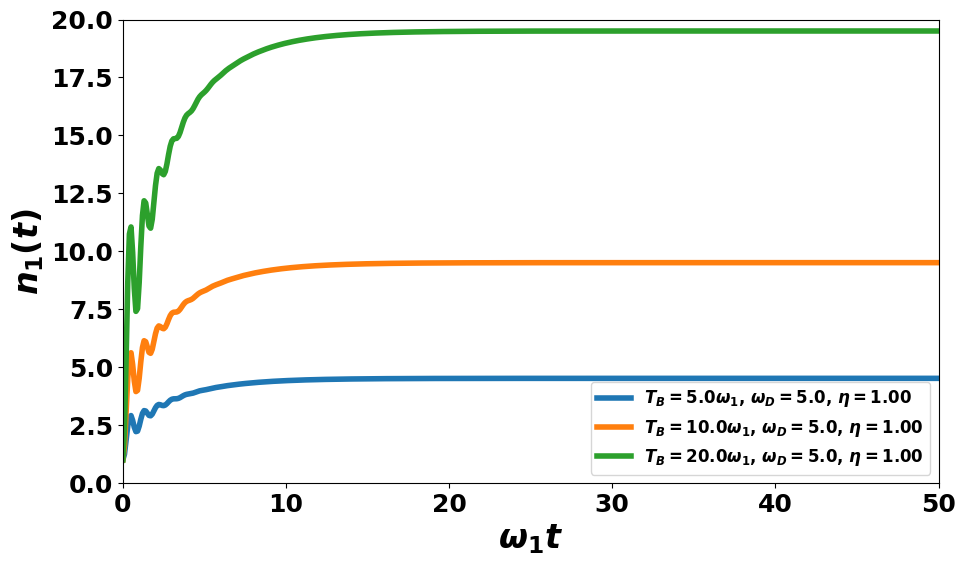}\caption{}\end{subfigure}\hfill
\begin{subfigure}{0.48\textwidth}
  \includegraphics[width=\textwidth]{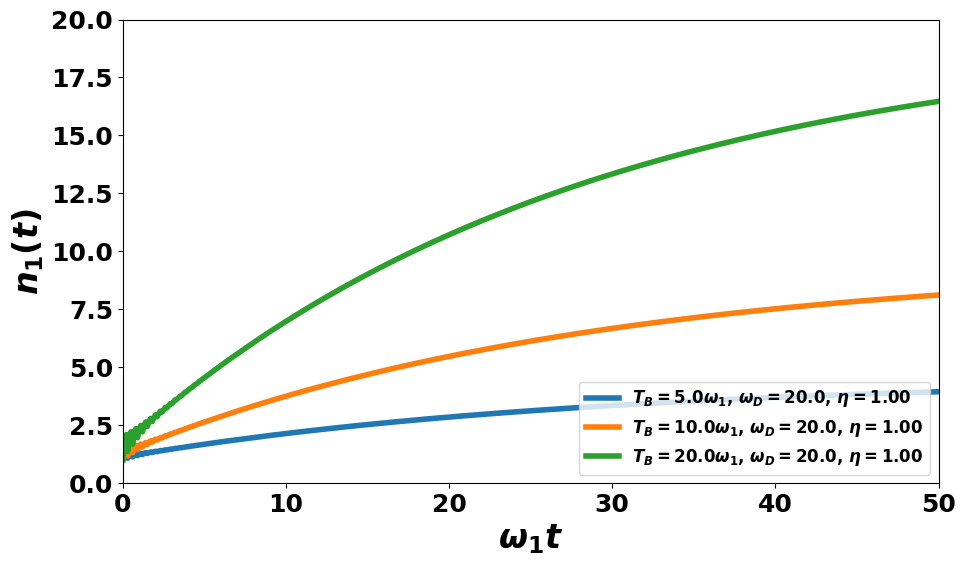}\caption{}\end{subfigure}\\[4pt]
\begin{subfigure}{0.48\textwidth}
  \includegraphics[width=\textwidth]{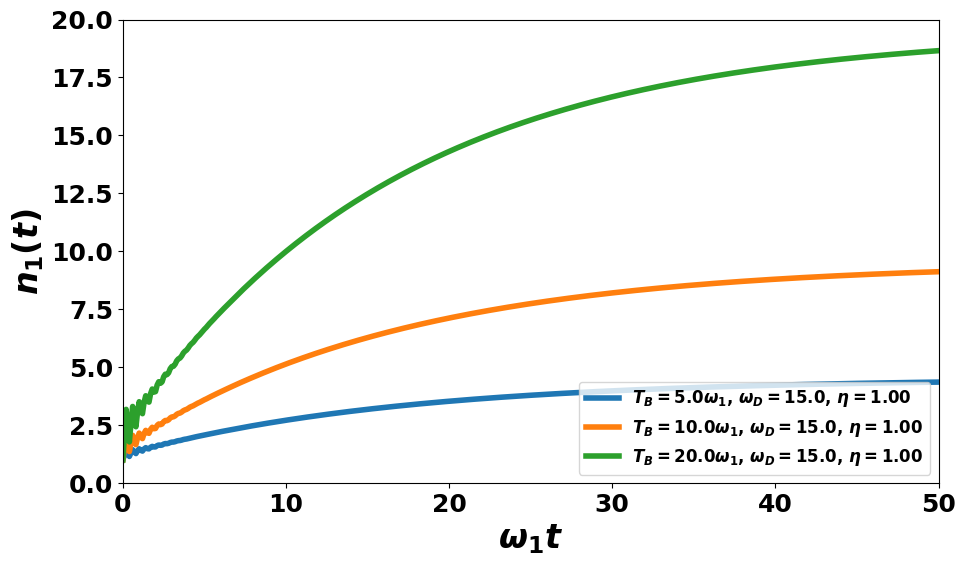}\caption{}\end{subfigure}\hfill
\begin{subfigure}{0.48\textwidth}
  \includegraphics[width=\textwidth]{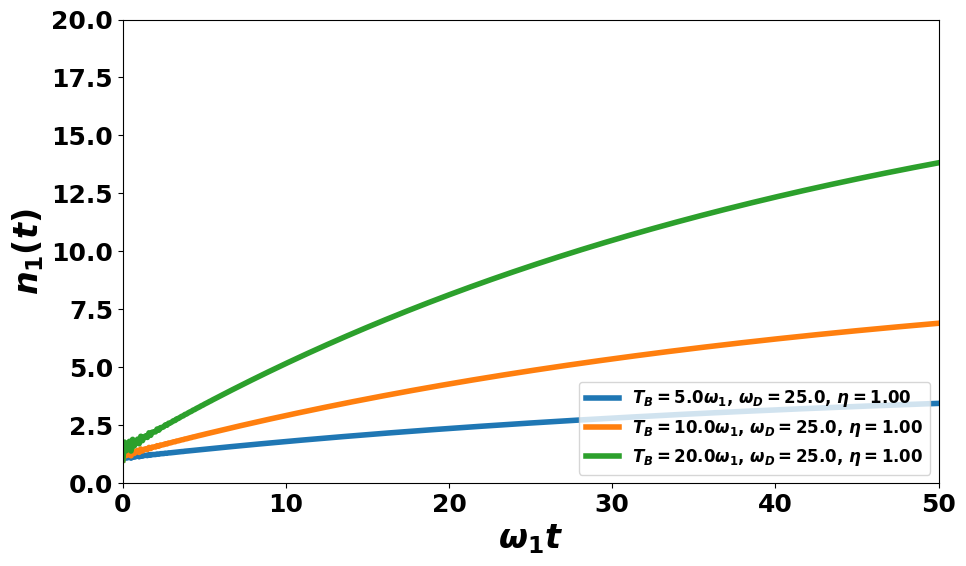}\caption{}\end{subfigure}
\caption{\textbf{Detuning amplitude at fixed duty cycle.} Time evolution of \(n_{1}(t)\) in the non-Markovian regime (\(\Gamma=15.0\), \(\Omega=1.0\), \(\omega_{1,0}=\omega_{2,0}=1.0\), \(\gamma=1.0\)) for \(\eta=1.0\) and increasing \(\omega_{D}\): (a) 5.0, (b) 15.0, (c) 20.0, (d) 25.0. Each panel shows three \(T_B\). Larger \(\omega_D\) drives the system further off resonance, suppressing non-Markovian revivals and lowering the steady-state AEN.}
\label{fig:fig7}
\end{figure*}

\begin{figure*}[t]
\centering
\begin{subfigure}{0.48\textwidth}
  \includegraphics[width=\textwidth]{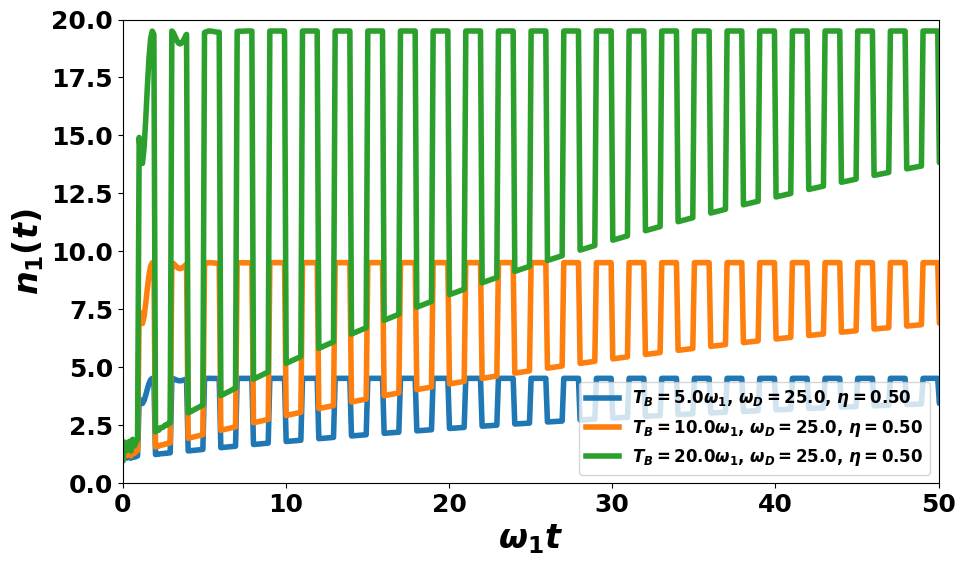}\caption{}\end{subfigure}\hfill
\begin{subfigure}{0.48\textwidth}
  \includegraphics[width=\textwidth]{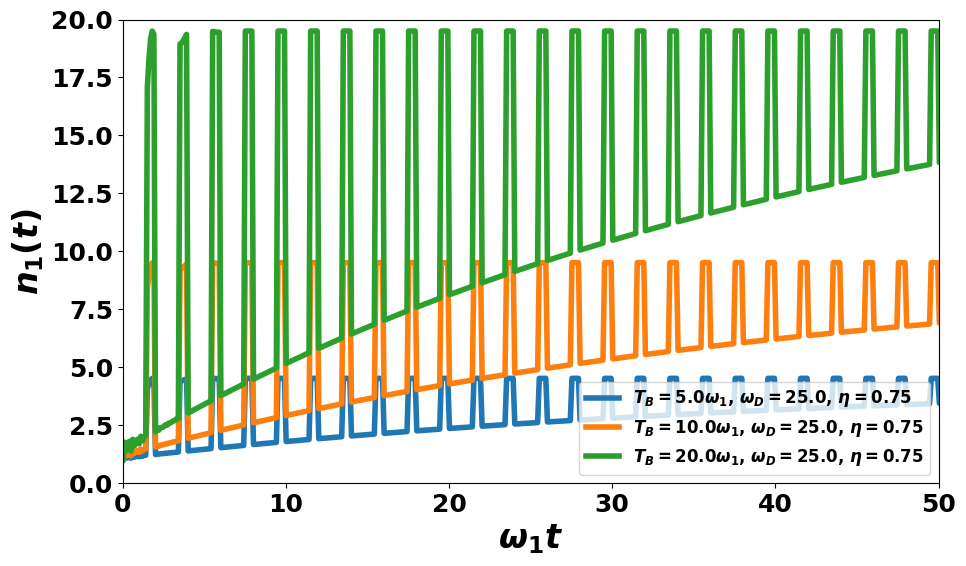}\caption{}\end{subfigure}\\[4pt]
\begin{subfigure}{0.48\textwidth}
  \includegraphics[width=\textwidth]{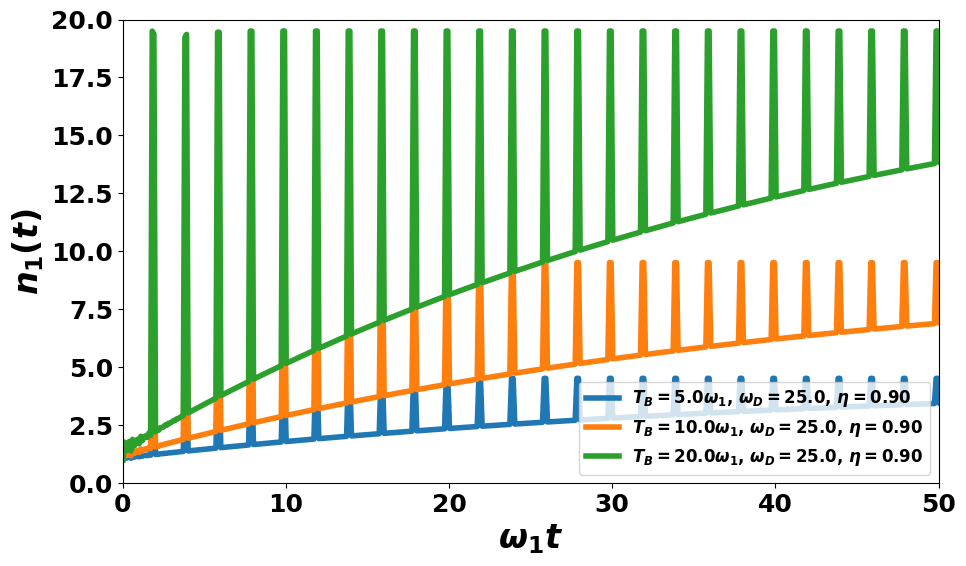}\caption{}\end{subfigure}\hfill
\begin{subfigure}{0.48\textwidth}
  \includegraphics[width=\textwidth]{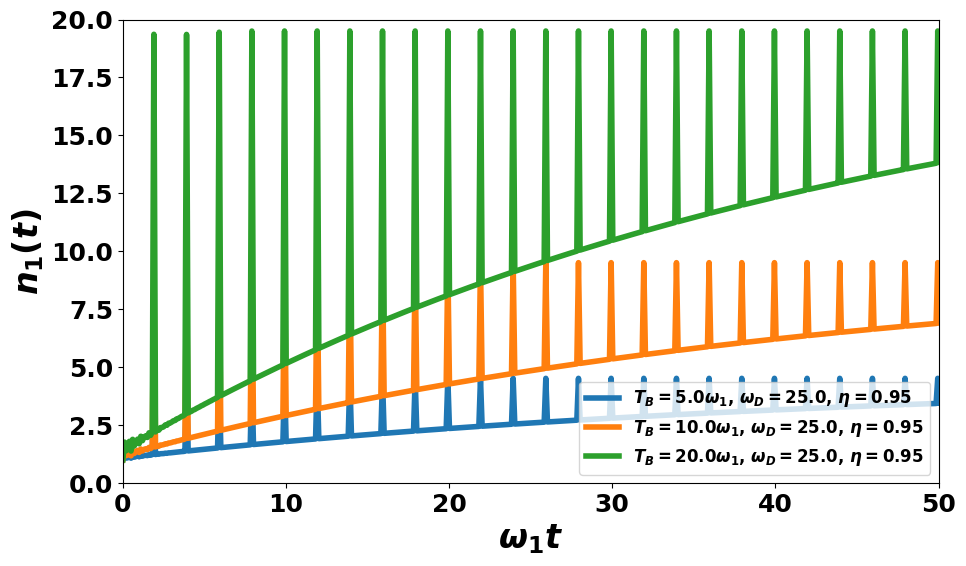}\caption{}\end{subfigure}
\caption{\textbf{Duty-cycle sweep at fixed detuning.} \(n_{1}(t)\) for \(\omega_D=25.0\) and \(\eta=\) (a) 0.50, (b) 0.75, (c) 0.90, (d) 0.95; same non-Markovian parameters as Fig.~\ref{fig:fig7}. Smaller \(\eta\) yields intermittent protection with visible steps; as \(\eta\to1\), modulation becomes effectively continuous and revivals are quenched.}
\label{fig:fig8}
\end{figure*}

Figures~\ref{fig:fig7} and~\ref{fig:fig8} extend the non-Markovian \emph{free evolution} shown in Fig.~\ref{fig:fig4} by activating our LEO-inspired, detuning-based DD. In the uncontrolled evolution ($\gamma=1.0$), the long bath memory produces oscillatory revivals in $n_1(t)$. Turning on DD shifts the system away from the bath’s spectral peak and reduces the spectral overlap, thereby damping these revivals and slowing the growth of the excitation number.

\paragraph*{Detuning amplitude at fixed duty cycle (\(\eta=1.0\)).}
Figure~\ref{fig:fig7} sweeps \(\omega_D\) while holding \(\eta=1.0\) (near-continuous modulation). Larger \(\omega_D\) pushes the system further off resonance, weakening system--bath exchange. Consequently, \(n_1(t)\) shows progressively stronger suppression of excitation growth and lower steady-state AEN, with non-Markovian ringing strongly damped. (For comparison, in the Markovian regime \(\Gamma=1.0,\gamma=15.0\), DD produces negligible changes, consistent with \cite{Jing2014}.)

\paragraph*{Duty cycle at fixed detuning amplitude.}
Figure~\ref{fig:fig8} fixes \(\omega_D=25.0\) and varies \(\eta\). For partial modulation (\(\eta<1\)), the curves exhibit step-like features: during OFF windows the system briefly re-enters near-resonance and absorbs energy. As \(\eta\) increases, protection lasts longer each period, revivals are quenched, and the approach to steady state becomes smoother; \(\eta\!\approx\!1\) closely matches the high-detuning, continuous case.

\begin{figure}[t]
\centering
\includegraphics[width=\linewidth]{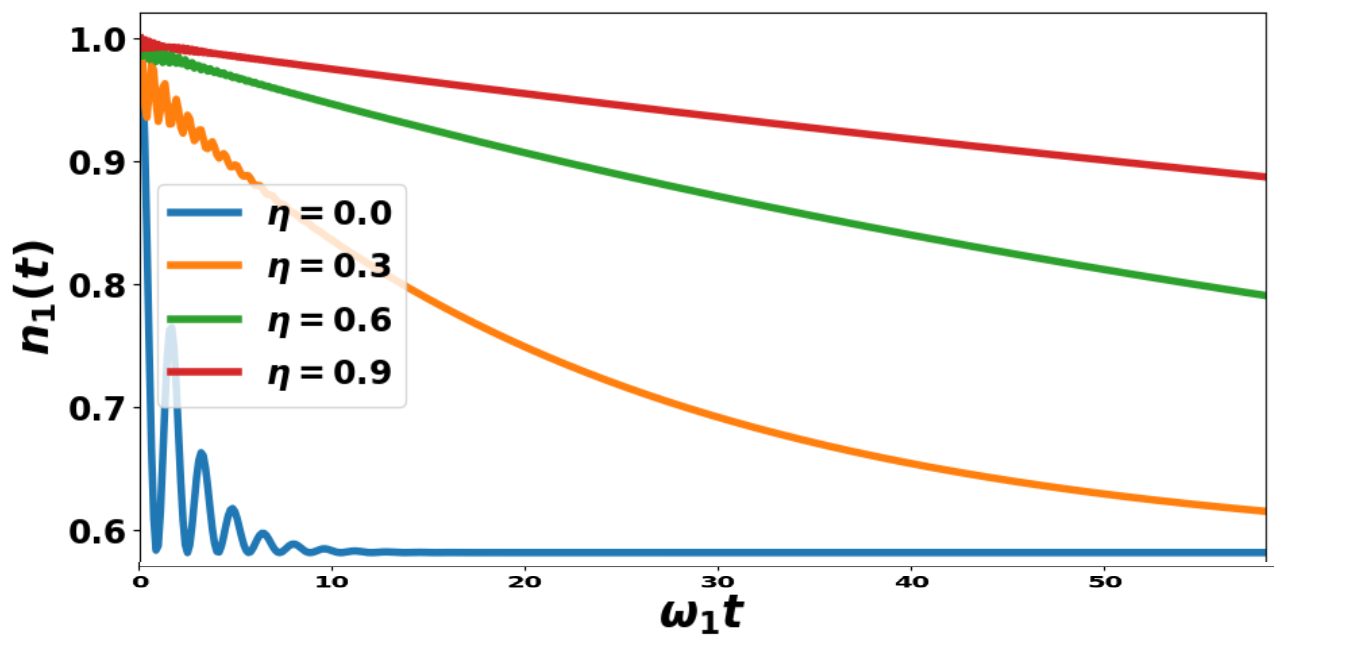}
\caption{\textbf{Isolating duty-cycle effects.} \(n_{1}(t)\) at fixed \(\omega_D=25.0\) and \(T_B=1.0\) while varying \(\eta\) from 0 to 0.9 (non-Markovian: \(\Gamma=15.0,\gamma=1.0,\Omega=\omega_{1,0}=\omega_{2,0}=1.0\)). Larger \(\eta\) reduces energy exchange and damps memory-induced oscillations; \(\eta\!\approx\!1\) gives maximal protection.}
\label{fig:fig9}
\end{figure}

\paragraph*{Duty-cycle–only scan at fixed \(T_B\).}
Figure~\ref{fig:fig9} isolates the role of \(\eta\) by fixing \(\omega_D=25.0\) and \(T_B=1.0\). Increasing \(\eta\) lengthens the protected fraction of each cycle, reduces net energy exchange, and damps memory-induced oscillations; small \(\eta\) leaves long free-evolution intervals where stored bath correlations re-excite the system.

\noindent
In long-memory baths, both knobs matter: larger \(\omega_D\) and higher \(\eta\) jointly suppress heating and non-Markovian revivals; in memoryless baths the effect is negligible~\cite{Jing2014}.

\begin{figure}[h!]
    \centering
    \includegraphics[width=1.0\linewidth]{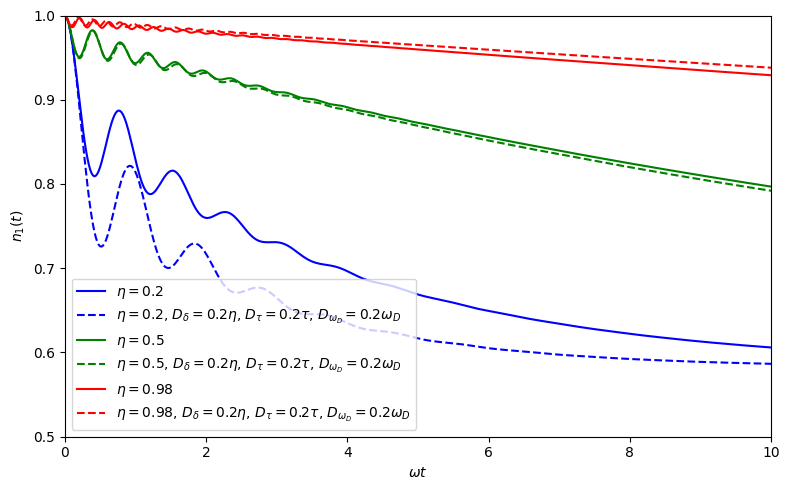}
    \caption{Time evolution of $n_{1}(t)$ under \emph{regular} (solid) and \emph{irregular} (dashed) dynamical–decoupling (DD) control in the non-Markovian regime.  
    System parameters: $\Gamma = 15.0$, $\Omega = 1.0$, $\omega_{1,0} = \omega_{2,0} = 1.0$, $\gamma = 1.0$, $\omega_{D} = 30.0$, and $T_{B} = 1.0$.  
    Duty cycles: $\eta = 0.2$ (blue), $0.5$ (green), and $0.98$ (red), where $\eta = \delta/\tau$.  
    In the irregular DD case, parameters are perturbed by $D_{\delta} = 0.2\delta$, $D_{\tau} = 0.2\tau$, and $D_{\omega_D} = 0.2\omega_D$, simulating realistic fluctuations in pulse width, period, and detuning amplitude.}
    \label{fig:fig10}
\end{figure}

Figure~\ref{fig:fig10} shows how $n_{1}(t)$ changes over time for \emph{regular} DD (solid lines) and \emph{irregular} DD (dashed lines) in a non-Markovian system.  
The main system parameters are: $\Gamma = 15.0$, $\Omega = 1.0$, $\omega_{1,0} = \omega_{2,0} = 1.0$, $\gamma = 1.0$, $\omega_{D} = 30.0$, and $T_{B} = 1.0$.  
We test three duty cycles: $\eta = 0.2$ (blue), $\eta = 0.5$ (green), and $\eta = 0.98$ (red), where $\eta = \delta / \tau$ is the fraction of each cycle that the control field is on.  

In the irregular DD case, we introduce random variations of $\pm 20\%$ to the pulse width $\delta$, the cycle time $\tau$, and the detuning amplitude $\omega_D$,  
so that $D_{\delta} = 0.2\delta$, $D_{\tau} = 0.2\tau$, and $D_{\omega_D} = 0.2\omega_D$.  
These variations model realistic imperfections in timing, width, and frequency, meaning that each parameter can be up to 20\% larger or smaller than its nominal value in each cycle.

For the small duty cycle $\eta = 0.2$, the control is active for only a short part of each cycle.  
The system interacts freely with the bath for most of the time, leading to fast decay of $n_{1}(t)$ and strong non-Markovian oscillations.  
Here, irregularity clearly worsens performance, allowing even more energy to leak away.

For the medium duty cycle $\eta = 0.5$, the control is on for half of each cycle, keeping the system off-resonance for longer and slowing decoherence.  
The difference between regular and irregular DD is much smaller than in the $\eta = 0.2$ case, but regular DD still performs slightly better.

For the large duty cycle $\eta = 0.98$, the control field is active almost all the time, keeping the system far off-resonance for nearly the entire evolution.  
Decoherence is strongly suppressed and $n_{1}(t)$ remains close to unity.  
In this regime, irregularity has almost no effect because the modulation is already strong and continuous-like.

In conclusion, increasing $\eta$ enhances decoherence suppression in non-Markovian baths. Irregularity has its greatest negative impact at small $\eta$, while for medium and large $\eta$ the effect is minimal.  
At $\eta \approx 1$, regular and irregular DD perform almost the same because the system is protected almost all the time.

\subsection{Control performance metric and evaluation procedure}
\label{sec:metrics}

To quantify how well our LEO-inspired, detuning–based DD suppresses
\emph{decoherence}, we compare each controlled trace directly to the
uncontrolled (``free'') evolution computed at the \emph{same} bath and
system parameters \((\Gamma,\gamma,\Omega,\omega_{1,0}=\omega_{2,0},T_B)\).

\paragraph*{Suppression factor.}
Following common practice in the filter-function framework and experimental DD benchmarks,
we quantify control by a single time-domain \emph{suppression factor} that normalizes the
observed decay under DD to the corresponding no–control decay~\cite{Cywinski2008,SuterAlvarez2016,Biercuk2009,deLange2010,Bylander2011,KhodjastehLidar2005,ViolaLloyd1998,ViolaKnillLloyd1999}:
\begin{equation}
S(t)=1-\frac{\big|\,n_1^{\mathrm{DD}}(t)-n_1^{\mathrm{free}}(0)\,\big|}
   {\big|\,n_1^{\mathrm{free}}(t)-n_1^{\mathrm{free}}(0)\,\big|}
\;\in(-\infty,1] .
\label{eq:eq28}
\end{equation}
If $S(t)=1$, this signifies \emph{perfect suppression} (the controlled observable stays at its starting level); $0<S(t)<1$ indicates \emph{partial suppression}; $S(t)=0$ means \emph{no improvement} over free evolution; and $S(t)<0$ indicates the control is \emph{detrimental} (rare in our parameter ranges).

To compute the suppression factor, the exact dynamics are solved twice: once without control to obtain the free trace $n_1^{\mathrm{free}}(t)$ and, for each DD setting (regular or irregular), once with control to obtain $n_1^{\mathrm{DD}}(t)$, using identical initial conditions and bath parameters in all cases. Both traces are sampled on the same time grid, and $S(t)$ is evaluated pointwise from Eq.~(\ref{eq:eq28}). To avoid numerical artifacts from a near-vanishing denominator at very early times, reporting begins after a short transient (e.g., $t\gtrsim 2/\Omega$), and small numerical overshoots are clipped so that $S(t)\le 1$. 

\begin{figure*}[t]
  \centering
  \begin{subfigure}{0.48\textwidth}
    \centering
    \includegraphics[width=\linewidth]{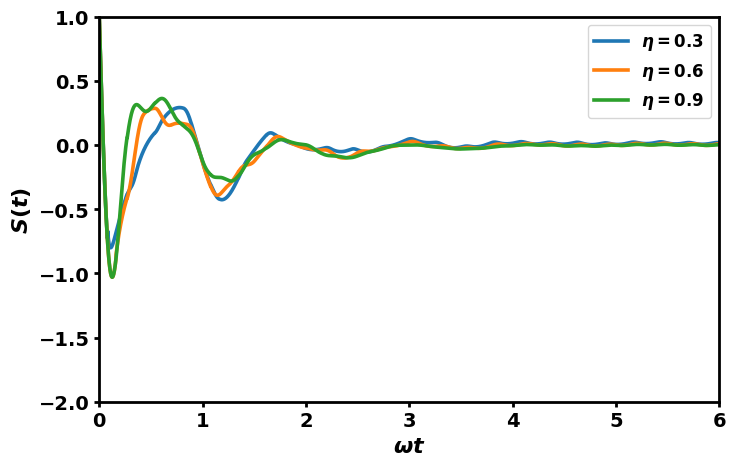}
    \caption{Regular DD: fixed detuning (e.g., $\omega_D=25$)
    with duty–cycle sweep; bath: $\Gamma=15$, $\gamma=1$, $\Omega=\omega_{1,0}=\omega_{2,0}=1$,
    $T_B=1$. Higher duty cycles yield $S(t)$ closer to~1 in the transient window.}
    \label{fig:fig11a}
  \end{subfigure}\hfill
  \begin{subfigure}{0.48\textwidth}
    \centering
    \includegraphics[width=\linewidth]{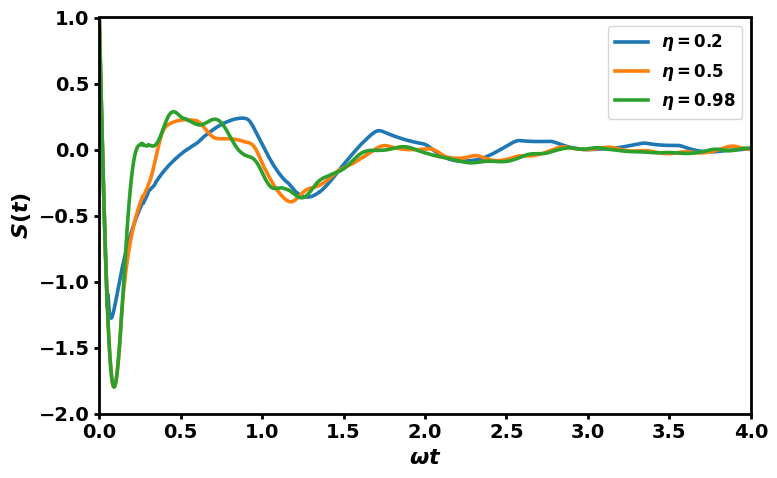}
    \caption{Irregular (jittered) DD: same bath and $T_B$, with $\pm 20\%$ jitter in
    $\delta$, $\tau$, and $\omega_D$ (e.g., $\omega_D=30$).
    At large duty cycle ($\eta\approx 1$) the irregular traces achieve $S(t)$
    comparable to regular DD; at small duty cycle suppression is weaker.}
    \label{fig:fig11b}
  \end{subfigure}
  \caption{\textbf{Suppression factor} $S(t)$ (Eq.~\ref{eq:eq28}) comparing DD vs.\ free
  evolution under identical bath parameters. $S(t)=1$ denotes perfect
  decoherence suppression; $S(t)=0$ denotes no improvement over free;
  $S(t)<0$ indicates transient degradation.}
  \label{fig:fig11}
\end{figure*}

Fig.~\ref{fig:fig11a} and Fig.~\ref{fig:fig11b} illustrate the \emph{suppression factor} $S(t)$ comparing a DD trajectory against the corresponding free (uncontrolled) trajectory under the \emph{same} bath and system parameters. Recall the interpretation: $S(t)=1$ means perfect suppression; $S(t)=0$ means no improvement over free evolution; and $S(t)<0$ means the DD trace deviates more than the free one (i.e., worse than free at that instant).

\paragraph*{Panel (a): Regular DD (duty–cycle sweep).}
We keep the detuning amplitude fixed and vary the duty cycle
$\eta\in\{0.3,0.6,0.9\}$ in a non-Markovian bath
($\Gamma=15$, $\gamma=1$, $\Omega=\omega_{1,0}=\omega_{2,0}=1$, $T_B=1$; e.g.,
$\omega_D=25$, $\tau=0.27$).
All curves share the same qualitative shape:
(i) an early dip below zero (a transient where switching and bath backflow can
make the DD trajectory move away from the initial value a bit faster than the
free one),
(ii) a positive lobe where DD \emph{does} suppress decoherence
($S(t)>0$), and then
(iii) a slow return toward $S(t)\to 0$ as both DD and free traces relax to the
\emph{same} thermal fixed point.
Larger duty cycles are better in the transient window:
the first negative dip is shallower and the positive lobe is slightly higher
for $\eta=0.9$ than for $\eta=0.3$, reflecting that the system spends more time
off resonance each period.

\paragraph*{Panel (b): Irregular DD (jittered pulses).}
We now introduce $\pm 20\%$ cycle-to-cycle jitter in pulse width $\delta$, period
$\tau$, and detuning amplitude $\omega_D$ (e.g., $\omega_D=30$), and sweep
$\eta\in\{0.2,0.5,0.98\}$ with the same bath as in panel~(a).
Irregular DD keeps the same overall behavior: a brief negative dip,
a positive lobe, then $S(t)\to 0$ at long times. Two trends are clear.
First, increasing $\eta$ again helps: the curve for $\eta=0.98$ has the
smallest initial dip and the largest positive lobe. Second, jitter mainly
hurts at small duty cycle: for $\eta=0.2$ the dip is deeper and the positive
lobe smaller, while for $\eta\approx 1$ the irregular traces are nearly as good
as regular ones. This matches DD “filter” intuition: randomizing the timing
smears the spectral notch, which matters little when protection is almost
continuous but can reduce peak suppression when the control is only on for a
short fraction of each cycle.

\paragraph*{Takeaway.}
These $S(t)$ plots quantify what we see in the raw occupations $n_1(t)$:
bigger duty cycles (and, elsewhere, bigger detuning) give better transient
suppression; irregular (randomized) DD is robust at high duty cycle but can be
slightly weaker at small duty cycle. Because both controlled and free
dynamics relax to the same thermal limit, $S(t)$ naturally drifts to $0$ at
long times—the metric evaluates \emph{how much decoherence is slowed down}, not a change of the final thermal state.

\section{Conclusions}
We have presented an exact, approximation–free analysis of two harmonic oscillators interacting with a common Lorentzian reservoir, and we have embedded open-loop control directly into this analytic framework. By eliminating the bath degrees of freedom and reducing the dynamics to a homogeneous second–order system, we solved the resulting characteristic quartic and constructed a probability-conserving propagator for the mode amplitudes. This gives closed–form expressions for the average excitation numbers and the inter–mode coherence, cleanly separating genuine environment–memory effects (revivals, backflow) from direct coherent exchange. Because the solution is exact, no Born or Markov assumptions are invoked, and the conserved norm ensures that all energy leaving the system is accounted for in the bath and can flow back according to the bath correlation time.

Furthermore, we implemented a leakage-elimination-operator (LEO)-inspired detuning control, treating both \emph{regular} (periodic) and \emph{irregular} (jittered) dynamical–decoupling (DD) pulse trains. The control enters as a piecewise constant frequency shift; each ON/OFF segment is propagated with the same closed–form homogeneous solution, and continuity of amplitudes and first derivatives is enforced at every switch. This yields a fully analytic \emph{piecewise–exact} integration of controlled non–Markovian dynamics without additional phenomenology.

\noindent Our main physical conclusions are as follows: in long–memory (non–Markovian) baths the uncontrolled system shows oscillatory revivals and transient backflow; in short–memory (effectively Markovian) baths relaxation is monotonic and revivals are quenched; detuning suppresses decoherence by repeatedly lifting resonance with the reservoir peak and reducing spectral overlap, so larger detuning amplitude \(\omega_D\) and higher duty cycle \(\eta\) more strongly curb heating and damp revivals, consistent with both the toggling–frame and filter–function pictures; comparing regular and irregular DD, cycle–to–cycle jitter primarily harms performance at small \(\eta\) by smearing spectral notches, while at high \(\eta\) the difference becomes negligible because protection is nearly continuous; a single time–domain suppression factor \(S(t)\) (pointwise comparison of the controlled trace to its free counterpart) captures the transient nature of DD, with \(S(t)\to 0\) at long times because both evolutions approach the same thermal fixed point; coherence is used here as a diagnostic rather than a resource—the exact propagator yields its time evolution, showing revivals in non–Markovian settings and strong damping under high–duty–cycle detuning, and in our parameter ranges it remains below standard separability bounds so none of the conclusions rely on entanglement generation. From these results, practical design rules follow without case–by–case tuning: pick a detuning amplitude large compared to the bath width to minimize spectral overlap during ON windows; maximize the duty cycle \(\eta\) within hardware limits since \(\eta\!\to\!1\) is the most robust and jitter–tolerant regime; choose a period \(\tau\) short compared to the bath correlation time \(1/\gamma\) because long OFF windows allow stored correlations to re–excite the system; and use \(S(t)\) to calibrate controls—the onset time, peak value, and time area where \(S(t)>0\) provide an operational measure of protection and indirectly reveal the bath memory scale. 

The present treatment has clear boundaries: it assumes identical system–bath couplings for the two modes, a Lorentzian spectral density, a common detuning applied to both modes, and resonance with the bath peak; we also focus on frequency–modulation DD rather than inversion (\(\pi\)–pulse) sequences and adopt a single–excitation initial condition for clarity. Looking ahead, the analytic machinery is portable: asymmetric couplings and detunings, alternative spectra (Ohmic, Drude–Lorentz, multi–peaked), and larger oscillator networks where bath–mediated transport competes with sparse coherent links are natural extensions; on the control side, shaping the detuning waveform beyond squares, hybridizing with inversion–based DD, and optimizing to maximize the time area with \(S(t)>0\) are promising directions; moreover, because \(S(t)\) and the coherence trace respond sensitively to \(\gamma\) and \(\Gamma\), the framework supports system–identification protocols that infer bath memory from transient suppression data. In sum, combining an exact probability–conserving propagator with a piecewise–exact treatment of detuning–based DD yields clear, quantitative rules for suppressing decoherence in long–memory environments and offers a rigorous benchmark for open–loop coherence protection in structured reservoirs.

\section*{Acknowledgement}
L.-A.W. is supported by the Basque Country Government (Grant No.\ IT1470-22) and Grant No.\ PGC2018-101355-B-I00 funded by MCIN/AEI/10.13039/501100011033, the Ministry for Digital Transformation and Civil Service of the Spanish Government through the QUANTUM ENIA project call—Quantum Spain project, and by the European Union through the Recovery, Transformation and Resilience Plan-NextGenerationEU within the framework of the Digital Spain 2026 Agenda.


\bibliography{main}
\appendix
\section{Derivation of the Heisenberg Equations}
\label{appendixA}

Starting from the total Hamiltonian in Eq.~\eqref{eq:eq1}, the Heisenberg equation of motion for any operator $\hat{O}$ is:
\begin{equation}
    \frac{d}{dt} \hat{O} = -i [\,\hat{O},\, H\,].
    \label{eq:A1}
\end{equation}
We take $\hbar = 1$ and $\partial\hat{O}/\partial t=0$ for the operators used here, so $\dot{\hat O} = i[H, \hat O]= -i[\hat O, H]$.\\

\noindent
\textbf{1) Mode $a_1$: Commutator structure}

For $a_1$ the total commutator separates as:
\[
[\, a_1,\, H\,] = [\, a_1,\, H_S\,] + [\, a_1,\, H_B\,] + [\, a_1,\, H_{SB}\,].
\]

The system Hamiltonian part expands as:
\[
H_S = \omega_1\, a_1^\dagger a_1 + \omega_2\, a_2^\dagger a_2 + g\, (\, a_1^\dagger a_2 + a_2^\dagger a_1 ).
\]

Standard bosonic commutators yield:
\[
\begin{aligned}
[\,a_1,\, H_S\,] &= \omega_1\, a_1 + g\, a_2,\quad 
&[\, a_1,\, H_B\,] &= 0, \\[6pt]
[\, a_1,\, H_{SB}\,] &= \sum_j \kappa_{1j} b_j.
\end{aligned}
\]

Therefore,
\begin{equation}
    \frac{d}{dt} a_1 = -i\, [\, a_1,\, H\,] 
    = -i\, \omega_1\, a_1 - i\, g\, a_2 - i \sum_j \kappa_{1j} b_j.
    \label{eq:A2}
\end{equation}

\noindent
\textbf{2) Modes $a_2$ and $b_j$}

Similarly,
\begin{align}
\frac{d}{dt} a_2 &= -i\, \omega_2\, a_2 - i\, g\, a_1 - i \sum_j \kappa_{2j} b_j,
\label{eq:A3} \\[6pt]
\frac{d}{dt} b_j &= -i\, \omega_j\, b_j - i\, \kappa_{1j} a_1 - i\, \kappa_{2j} a_2.
\label{eq:A4}
\end{align}

\noindent
Combining these gives Eq.~\eqref{eq:eq5} of the main text.

\paragraph*{Notes.}
(i) The sign convention in Eq.~\eqref{eq:A1} is used consistently throughout, so the signs in Eqs.~\eqref{eq:A2}–\eqref{eq:A4} follow directly from $-i[\,\hat O,H\,]$. (ii) The result $[\,a_1,H_B\,]=0$ reflects that $H_B$ acts solely on bath operators. (iii) The $H_{SB}$ commutator gives the exchange terms because of the rotating–wave form of $H_{SB}$.

\section{ Exact Quartic Solution and the Role of the Resolvent Cubic}
\label{AppendixB}

\noindent
\textbf{1) Coupled-mode system}

The homogeneous dynamics for the coupled amplitudes \( A_1(t) \) and \( A_2(t) \) are:
\begin{align}
\ddot{A}_1 + \alpha\, \dot{A}_1 + \beta\, A_1 + \gamma\, A_2 &= 0, \label{eq:B1} \\[6pt]
\ddot{A}_2 + \tilde{\alpha}\, \dot{A}_2 + \tilde{\beta}\, A_2 + \tilde{\gamma}\, A_1 &= 0. \label{eq:B2}
\end{align}

If the system is fully symmetric, then the cross-coupling is simply $\tilde{\gamma} = \gamma$.  
If the system parameters are asymmetric (e.g., $\omega_1 \neq \omega_2$), then the general form $\tilde{\gamma}$ must be used.

\noindent
\textbf{2) Exponential ansatz}

Assume:
\[
A_1(t) = X\, e^{\lambda t}, 
\quad 
A_2(t) = Y\, e^{\lambda t}.
\]

Substituting into Eqs.~(\ref{eq:B1})--(\ref{eq:B2}) gives:
\[
\begin{bmatrix}
\lambda^2 + \alpha\, \lambda + \beta & \gamma \\
\tilde{\gamma} & \lambda^2 + \tilde{\alpha}\, \lambda + \tilde{\beta}
\end{bmatrix}
\begin{bmatrix}
X \\ Y
\end{bmatrix}
= 0.
\]

The nontrivial solution condition is:
\[
\begin{aligned}
\det
\begin{bmatrix}
\lambda^2 + \alpha\, \lambda + \beta & \gamma \\[6pt]
\tilde{\gamma} & \lambda^2 + \tilde{\alpha}\, \lambda + \tilde{\beta}
\end{bmatrix}
&= 0, \\[6pt]
\Longrightarrow\;
\lambda^4 + A\, \lambda^3 + B\, \lambda^2 + C\, \lambda + D &= 0.
\end{aligned}
\]

with:
\[
A = \alpha + \tilde{\alpha}, 
\quad 
B = \alpha\, \tilde{\alpha} + \beta + \tilde{\beta}, 
\quad 
C = \alpha\, \tilde{\beta} + \tilde{\alpha}\, \beta, 
\quad 
D = \beta\, \tilde{\beta} - \gamma\, \tilde{\gamma}.
\]

\noindent
\textbf{3) Mode ratio}

When we solve the coupled ODE system, the amplitudes $X$ and $Y$ must satisfy a specific relation for each possible eigenvalue $\lambda_k$.

To see this, we insert the exponential ansatz $A_1(t) = X e^{\lambda t}$ and $A_2(t) = Y e^{\lambda t}$ into the first full ODE:
\[
\ddot{A}_1 + \alpha\, \dot{A}_1 + \beta\, A_1 + \gamma\, A_2 = 0.
\]

The derivatives yield:
\[
\dot{A}_1 = \lambda X e^{\lambda t}, 
\quad 
\ddot{A}_1 = \lambda^2 X e^{\lambda t}.
\]

Substituting these gives:
\[
\lambda^2 X\, e^{\lambda t} + \alpha\, \lambda X\, e^{\lambda t} + \beta X\, e^{\lambda t} + \gamma Y\, e^{\lambda t} = 0.
\]

Dividing by the common exponential factor $e^{\lambda t}$ leaves the algebraic condition:
\[
\big( \lambda_k^2 + \alpha\, \lambda_k + \beta \big) X + \gamma\, Y = 0.
\]

So, for each allowed eigenvalue $\lambda_k$, the two mode amplitudes must have the fixed ratio:
\[
r_k = \frac{Y}{X} = -\, \frac{ \lambda_k^2 + \alpha\, \lambda_k + \beta }{\, \gamma }.
\]

In simple terms, this means that for each collective oscillation mode, the strength of mode~2 is set by the motion of mode~1 through the direct coupling $\gamma$.  
The factor $r_k$ tells us exactly how much mode~2 responds when mode~1 oscillates at a given $\lambda_k$.

\noindent
\textbf{4) General solution}

Combining the mode ratio condition with the fact that the system supports four independent eigenmodes, the total solution is a sum over all four allowed pairs. 

Each eigenmode has the same time dependence for both amplitudes but is weighted by its specific mode ratio $r_k$:
\begin{align}
A_1(t) &= \sum_{k=1}^4 C_k\, e^{\lambda_k t}, \notag \\[6pt]
A_2(t) &= \sum_{k=1}^4 r_k\, C_k\, e^{\lambda_k t}.
\label{eq:B3}
\end{align}

This guarantees that the coupled equations, the characteristic quartic, and the eigenvector relation are all exactly satisfied.

\noindent
\textbf{5) Solving the quartic by Ferrari’s method}

To find the exact eigenvalues \(\lambda_k\), we solve the quartic characteristic equation:
\[
\lambda^4 + A\, \lambda^3 + B\, \lambda^2 + C\, \lambda + D = 0.
\]

Quartic equations can be solved in closed form, but the standard strategy is to first remove the cubic term by a simple change of variable:
\[
\lambda = y - \frac{A}{4}.
\]

Substituting this shift into the quartic cancels the \(y^3\) term and leaves a \emph{depressed quartic}:
\begin{align}
y^4 + p\, y^2 + q\, y + r &= 0, \label{eq:B4}
\end{align}
where, $ p = B - \frac{3}{8} A^2$, $q = C - \frac{1}{2} A B + \frac{1}{8} A^3$, $r = D - \frac{1}{4} A C + \frac{1}{16} A^2 B - \frac{3}{256} A^4$.
This form is simpler to handle because it has no cubic term. Ferrari’s method \cite{stewart2004galois, artin2010algebra} then factors this depressed quartic into a product of two quadratics by introducing an auxiliary parameter, which is determined by solving a related \emph{resolvent cubic}. Solving that resolvent cubic gives all the information needed to split the quartic exactly and find its four roots.

This method provides a fully analytic solution for the quartic eigenvalues \(\lambda_k\), which in turn fully determine the mode amplitudes and the system’s time evolution.

\noindent
\textbf{6) Factorization and resolvent cubic}

Ferrari’s factorization:
\[
y^4 + p\, y^2 + q\, y + r = (y^2 + a\, y + m)(y^2 - a\, y + n),
\]
leads to the resolvent cubic:
\begin{equation}
z^3 - \frac{p}{2} z^2 - r\, z + \Big( \frac{rp}{2} - \frac{q^2}{8} \Big) = 0.
\label{eq:B5}
\end{equation}

A real root \( z_0 \) gives:
\[
a = \sqrt{2 z_0 - p}, 
\quad 
b = -\, \frac{q}{2a}.
\]

\noindent
\textbf{Exact closed solution for the resolvent cubic Eq. \ref{eq:B5}:}

\noindent
Eq. \ref{eq:B5} can be written as:
\[
z^3 + A_c z^2 + B_c z + C_c = 0,
\quad 
\]
where $A_c = -\frac{p}{2}$, $B_c = -r$, $C_c = \frac{rp}{2} - \frac{q^2}{8}$.

\vspace{0.5em}

The exact solution for a real root is:
\[
z_0 = u + v - \frac{A_c}{3},
\quad 
\]
where $u = \sqrt[3]{ -\frac{Q}{2} + \sqrt{ \Big( \frac{Q}{2} \Big)^2 + \Big( \frac{P}{3} \Big)^3 } }$, $v = \sqrt[3]{ -\frac{Q}{2} - \sqrt{ \Big( \frac{Q}{2} \Big)^2 + \Big( \frac{P}{3} \Big)^3 } }$, $P = B_c - \frac{A_c^2}{3}$, $Q = 2\, \frac{A_c^3}{27} - \frac{A_c B_c}{3} + C_c$.

Once $z_0$ is known, the key auxiliary factor is:
\[
a = \sqrt{\, 2 z_0 - p\, },
\quad 
b = -\, \frac{q}{2a}.
\]

\vspace{0.5em}

\noindent
\textbf{Remark:} These coefficients $p, q, r$ depend on the system-bath parameters through 
$A, B, C, D$, which in turn depend on $ \alpha, \tilde{\alpha}, \beta, \tilde{\beta}, \gamma$, and $\tilde{\gamma}$.  
Their explicit forms are:
\[
\begin{cases}
A = \alpha + \tilde{\alpha}, \\[6pt]
B = \alpha \tilde{\alpha} + \beta + \tilde{\beta}, \\[6pt]
C = \alpha \tilde{\beta} + \tilde{\alpha} \beta, \\[6pt]
D = \beta \tilde{\beta} - \gamma \tilde{\gamma}.
\end{cases}
\quad
\text{and}
\quad
\begin{cases}
\alpha = \gamma + i(\Omega + \omega_1), \\[6pt]
\tilde{\alpha} = \gamma + i(\Omega + \omega_2), \\[6pt]
\beta = \frac{\Gamma \gamma}{2} - \omega_1 \Omega + i \gamma \omega_1, \\[6pt]
\tilde{\beta} = \frac{\Gamma \gamma}{2} - \omega_2 \Omega + i \gamma \omega_2.
\end{cases}
\]

Thus the resolvent cubic closes the Ferrari solution for the quartic.  
\textbf{This completes the explicit analytical route to obtain all quartic roots in radicals.}

Roots:
\begin{equation}
\begin{cases}
y_{1,2} = +\, \frac{a}{2} \pm \sqrt{\, z_0 - b\, }, \\[6pt]
y_{3,4} = -\, \frac{a}{2} \pm \sqrt{\, z_0 + b\, }.
\end{cases}
\quad 
\lambda = y - \frac{A}{4}.
\label{eq:B6}
\end{equation}

\noindent
\textbf{7) System-specific form}

Roots \(\lambda_k\) are fully determined by the physical parameters:
\(\gamma, \Gamma, g, \Omega, \omega_1, \omega_2\),
via \(\alpha, \tilde{\alpha}, \beta, \tilde{\beta}, \gamma, \tilde{\gamma}\).

\begin{widetext}

\begin{align}
\lambda_{1,2} &= -\,\frac{\gamma}{2} - \frac{i}{4}(2\Omega + \omega_1 + \omega_2)
+\, \frac{1}{2} \sqrt{
\, 2 z_0 - 
\Big[\, \gamma^2 + \Gamma\, \gamma - (\Omega + \omega_1)(\Omega + \omega_2) 
+ i\, \gamma\, [\, 2\Omega + 2(\omega_1 + \omega_2) ] \Big]
- \frac{3}{8} \big[\, 2\gamma + i(2\Omega + \omega_1 + \omega_2) \big]^2
} \notag \\[6pt]
&\quad 
\pm\, \frac{1}{2} \sqrt{
\, -\, 2 z_0 - 
\Big[\, \gamma^2 + \Gamma\, \gamma + i\, \gamma\, [\, 2\Omega + 2(\omega_1 + \omega_2) ] 
- \Omega^2 - 2\Omega(\omega_1 + \omega_2) - \omega_1 \omega_2 \Big]
- \frac{3}{4} \big[\, 2\gamma + i(2\Omega + \omega_1 + \omega_2) \big]^2
 - 2b },
\label{eq:B7}
\end{align}

\begin{align}
\lambda_{3,4} &= -\,\frac{\gamma}{2} - \frac{i}{4}(2\Omega + \omega_1 + \omega_2)
-\, \frac{1}{2} \sqrt{
\, 2 z_0 - 
\Big[\, \gamma^2 + \Gamma\, \gamma - (\Omega + \omega_1)(\Omega + \omega_2) 
+ i\, \gamma\, [\, 2\Omega + 2(\omega_1 + \omega_2) ] \Big]
- \frac{3}{8} \big[\, 2\gamma + i(2\Omega + \omega_1 + \omega_2) \big]^2
} \notag \\[6pt]
&\quad 
\pm\, \frac{1}{2} \sqrt{
\, -\, 2 z_0 - 
\Big[\, \gamma^2 + \Gamma\, \gamma + i\, \gamma\, [\, 2\Omega + 2(\omega_1 + \omega_2) ] 
- \Omega^2 - 2\Omega(\omega_1 + \omega_2) - \omega_1 \omega_2 \Big]
- \frac{3}{4} \big[\, 2\gamma + i(2\Omega + \omega_1 + \omega_2) \big]^2
 - 2b }.
\label{eq:B8}
\end{align}

\noindent
Here, the auxiliary terms are:
\[
b = 
\Big(
\gamma + i(\Omega + \omega_1)
\Big)
\Big(
\tfrac{\Gamma \gamma}{2} - \omega_2 \Omega + i\, \gamma\, \omega_2
\Big)
+ 
\Big(
\gamma + i(\Omega + \omega_2)
\Big)
\Big(
\tfrac{\Gamma \gamma}{2} - \omega_1 \Omega + i\, \gamma\, \omega_1
\Big),
\quad
z_0 = \text{real solution of the resolvent cubic}.
\]

\vspace{1em}

\end{widetext}

\noindent
\textbf{8) Initial conditions and final amplitudes}

To fully determine the solution, the constants $C_k$ are fixed by the initial conditions.  
We impose:
\[
A_1(0) = 1, \quad A_2(0) = 0, 
\quad \dot{A}_1(0) = -\, i\, \omega_1, 
\quad \dot{A}_2(0) = -\, i\, g.
\]

Substituting the general solutions,
\[
A_1(t) = \sum_{k=1}^4 C_k\, e^{\lambda_k t}, 
\quad
A_2(t) = \sum_{k=1}^4 r_k\, C_k\, e^{\lambda_k t},
\]

we get:
\[
A_1(0) = \sum_{k=1}^4 C_k = 1,
\quad
A_2(0) = \sum_{k=1}^4 r_k\, C_k = 0.
\]

For the time derivatives,
\[
\dot{A}_1(t) = \sum_{k=1}^4 \lambda_k\, C_k\, e^{\lambda_k t},
\quad
\dot{A}_2(t) = \sum_{k=1}^4 \lambda_k\, r_k\, C_k\, e^{\lambda_k t}.
\]

Evaluating at $t=0$:
\[
\dot{A}_1(0) = \sum_{k=1}^4 \lambda_k\, C_k = -\, i\, \omega_1,
\quad
\dot{A}_2(0) = \sum_{k=1}^4 \lambda_k\, r_k\, C_k = -\, i\, g.
\]

So the four conditions are:
\[
\begin{cases}
C_1 + C_2 + C_3 + C_4 = 1, \\[6pt]
r_1\, C_1 + r_2\, C_2 + r_3\, C_3 + r_4\, C_4 = 0, \\[6pt]
\lambda_1\, C_1 + \lambda_2\, C_2 + \lambda_3\, C_3 + \lambda_4\, C_4 = -\, i\, \omega_1, \\[6pt]
\lambda_1\, r_1\, C_1 + \lambda_2\, r_2\, C_2 + \lambda_3\, r_3\, C_3 + \lambda_4\, r_4\, C_4 = -\, i\, g.
\end{cases}
\]

\noindent
Putting this in compact \textbf{matrix form}:
\[
M\, \mathbf{C} = 
\begin{bmatrix}
1 \\ 0 \\ -\, i\, \omega_1 \\ -\, i\, g
\end{bmatrix},
\quad 
M =
\begin{bmatrix}
1 & 1 & 1 & 1 \\[6pt]
r_1 & r_2 & r_3 & r_4 \\[6pt]
\lambda_1 & \lambda_2 & \lambda_3 & \lambda_4 \\[6pt]
\lambda_1 r_1 & \lambda_2 r_2 & \lambda_3 r_3 & \lambda_4 r_4
\end{bmatrix}.
\quad 
\mathbf{C} =
\begin{bmatrix}
C_1 \\ C_2 \\ C_3 \\ C_4
\end{bmatrix}.
\]

\noindent
Solving this linear system gives all the $C_k$, which then fully specify $A_1(t)$ and $A_2(t)$.

---

\noindent
\textbf{9) Final population expressions}

\begin{align}
n_1(t) &= |A_1(t)|^2\, n_{10} + |A_2(t)|^2\, n_{20} + \big[\, 1 - |A_1(t)|^2 - |A_2(t)|^2 \big]\, n_B, 
\label{eq:B9} \\[6pt]
n_2(t) &= |A_2(t)|^2\, n_{10} + |A_1(t)|^2\, n_{20} + \big[\, 1 - |A_1(t)|^2 - |A_2(t)|^2 \big]\, n_B.
\label{eq:B10}
\end{align}

(i) For numerical stability, choose square–root branches in Eqs.~\eqref{eq:B7}–\eqref{eq:B8} consistently across time steps (e.g., by continuity in $\lambda_k$). (ii) Sorting modes by $\mathrm{Re}\,\lambda_k$ is helpful when enforcing decaying transients. (iii) The matrix system for $\mathbf{C}$ is well-conditioned if the $\lambda_k$ are nondegenerate; near degeneracies, regularization or high-precision arithmetic may be useful.

\noindent
This completes the exact quartic solution with proper cross-coupling.

\end{document}